\documentclass[a4paper]{amsart}
\usepackage{amscd,amsmath,amsthm,amssymb,marvosym}
\usepackage[utf8]{inputenc}
\usepackage{color}
\usepackage[english]{babel}
\usepackage{graphicx,wrapfig}
\usepackage{sub caption}
\usepackage[all]{xy}
\usepackage{soul}
\usepackage{pgf,tikz}
\usepackage[colorlinks=true, linkcolor=blue, citecolor=blue]{hyperref}
\usepackage{mathtools}
\usetikzlibrary{shapes.geometric,patterns,arrows,decorations.markings,decorations.pathmorphing}
\usepackage{animate}

\def\red{\color{red}}

\def\black{\color{black}}
\def\white{\color{white}}
\definecolor{cambridge}{RGB}{174,0,0}
\definecolor{darkblue}{RGB}{0,0,174}
\definecolor{darkgreen}{RGB}{0,174,0}
\definecolor{darkgrey}{RGB}{50,50,50}

\def\a{\alpha}
\def\b{\beta}
\def\ga{\gamma}
\def\de{\delta}   

\def\la{\lambda}

\def\r{\rho}
\def\s{\sigma}
\def\z{\zeta}
\def\om{\omega}
\def\Om{\Omega}

\def\C{{\bb C}}
\def\E{{\mathbb E}}

\def\N{{\bb N}}

\def\bbP{{\bb P}}
\def\Q{{\bb Q}}
\def\R{{\bb R}}
\def\S{{\bb S}}

\def\Z{{\bb Z}}

\def\bb#1{\mathbb{#1}}
\def\bf#1{\mathbf{#1}}

\def\sC{{\mathcal C}}

\def\sN{{\mathcal N}}

\def\sR{{\mathcal R}}

\def\sZ{{\mathcal Z}}
\def\Ga{\Gamma}

\def\La{\Lambda}

\def\x{\times}

\def\o+{\oplus}

\def\<{\langle}
\def\>{\rangle}

\def\({\left(}
\def\){\right)}
\def\=#1{\bar #1}
\def\~#1{\widetilde #1}

\def\.#1{\dot #1}
\def\^#1{\widehat #1}
\def\bra{\langle}
\def\ket{\rangle}

\def\eeq{\end{split}\end{equation}}
\def\beq{\begin{equation}\begin{split}}

\def\beql#1{\begin{equation} \label{#1}\begin{split}}

\def\eqref#1{(\ref{#1})}

\newtheorem{theorem}{Theorem}[section]

\newtheorem{lemma}[theorem]{Lemma}
\newtheorem{proposition}[theorem]{Proposition}

\newtheorem{question}{Question}[section]
\def\bask{\red\begin{question}}
\def\cask{\end{question}\black}

\newtheorem{Ex}{Ex}

\theoremstyle{definition}
\newtheorem{definition}{Definition}[section]

\def\corner{\Om^{\diamond}}
\def\midpoint{\Om^{\flat}}
\def\dual{\Om^{*}}

\newcommand\cle{\operatorname{CLE}}
\newcommand\pfaffian{\operatorname{Pfaff}}

\newcommand\sign{\operatorname{sign}}

\renewcommand\Re{\operatorname{Re}}

\def\sec#1{\section{#1}\label{#1}}
\def\subsec#1{\subsection{#1}\label{#1}}
\def\secref#1{Section \ref{#1}}

\def\figref#1{Figure \ref{#1}}

\begin{document}


\newdimen\pgfdecorationcornersradius\newdimen\pgfdecorationborderwidth
\newdimen\a\newdimen\b\newdimen\o\newdimen\R\newdimen\r
\pgfkeys{
  /pgf/decoration/.cd,
  rounded corners/.code={\pgfmathsetlength\pgfdecorationcornersradius{#1}},
  border width/.code={\pgfmathsetlength\pgfdecorationborderwidth{#1}},
  fill/.initial=black
}

\pgfdeclaredecoration{dumbbells}{initial}
{
  \state{initial}[width=0,next state=depict one]{
    \egroup
    \pgfsetfillcolor{\pgfkeysvalueof{/pgf/decoration/fill}}
    \pgfmathsetlength\a{.5\pgfdecorationsegmentlength}
    \pgfmathsetlength\b{1.41421\pgfdecorationborderwidth}
    \pgfmathsetlength\R{\pgfdecorationcornersradius+\pgfdecorationborderwidth}
    \pgfmathsetlength\r{\pgfdecorationcornersradius-\pgfdecorationborderwidth}
    \bgroup
  }
  \state{depict one}[width=\pgfdecorationsegmentlength,next state=depict one]
  {
    \pgfpathmoveto{\pgfpoint{\o}{\a+\b}}
    \pgfsetcornersarced{\pgfpoint{\r}{\r}}
    \pgfpathlineto{\pgfpoint{\a}{\b}}
    \pgfsetcornersarced{\pgfpoint{\R}{\R}}
    \pgfpathlineto{\pgfpoint{2\a}{\a+\b}}
    \pgfpathlineto{\pgfpoint{3\a+\b}{\o}}
    \pgfpathlineto{\pgfpoint{2\a}{-\a-\b}}
    \pgfsetcornersarced{\pgfpoint{\r}{\r}}
    \pgfpathlineto{\pgfpoint{\a}{-\b}}
    \pgfsetcornersarced{\pgfpoint{\R}{\R}}
    \pgfpathlineto{\pgfpoint{\o}{-\a-\b}}
    \pgfpathlineto{\pgfpoint{-\a-\b}{\o}}
    \pgfpathclose
    \pgfusepath{fill}
    \pgfpathmoveto{\pgfpointorigin}
  }
  \state{final}{}
}
\tikzset{
    fill dumbbells/.style 2 args={
        decorate,decoration={
            dumbbells,
            segment length=1cm,
            rounded corners=.2cm,
            border width=#1,
            fill=#2
        }
    },
    draw dumbbells/.style={
        preaction={
            preaction={fill dumbbells={.01cm}{black}},
            postaction={fill dumbbells={-.01cm}{white}}
        },
        draw,
    }
}
\title{On the construction of discrete fermions in the FK-Ising model}
\author[Francesco Spadaro]{Francesco Spadaro}\thanks{\'Ecole Polytechnique F\'ed\'erale de Lausanne,  Switzerland; {\tt francesco.spadaro@epfl.ch}}
\maketitle
\begin{abstract}
We consider many-point correlation functions of discrete fermions in the two-dimensional FK-Ising model and show that, despite not being commuting observable, they can be realized with a geometric-probabilistic approach in terms of loops of the model and their winding.
\end{abstract}

\section{Introduction}
In the context of two-dimensional Statistical Field Theory, by exploiting conformal symmetry, big achievements have been made in comprehending the structure of the collection of different fields arising in statistical mechanics models at criticality.
From a statistical mechanics point of view --where the interest is in connecting discrete probabilistic models with their continuum scaling limits-- correlation functions of {\it bosonic fields} at the continuum level should be understood as scaling limits of expectations of discrete random variables in precursor models. Consider, for instance, the well celebrated Ising model at criticality on a discretization $\Om_{\de}$ of a simply connected domain $\Om$. In \cite{chelkak2015conformal} it has been established that the (properly rescaled) average of spin products converges, in the mesh size limit $\de\downarrow0$, to correlation functions of the spin field in the Ising Conformal Field Theory 
$$\de^{-\frac{n}{8}}{\bb E}_{\Om_{\de}}[\sigma_{z_{1}^{\de}}\dots\sigma_{z_{n}^{\de}}]\xrightarrow{\de\to0}{\sC}^{n}\langle\sigma_{z_{1}}\dots\sigma_{z_{n}}\rangle_{\Om}\ .$$

Being limits of averages of scalar real random variables, these correlation functions commute with respect to permutations of the order of the insertion points $z_{1},\dots,z_{n}$.

However, the study of continuum Conformal Field Theories (CFT) has been extended to non-bosonic fields --i.e. fields with non-commuting correlation functions with respect to exchanging insertion points--; the most famous example is the description of the Ising CFT as a free fermionic field theory: here a pivotal role is played by the fermionic field $\psi$ \cite{mussardo2010statistical,francesco2012conformal,henkel2013conformal}. Unlike bosonic fields, correlation functions of fermions do not commute but rather anticommute, i.e. when permuting the insertion points the correlation function gains a sign equal to the sign of the permutation. This behavior makes the interpretation of CFT correlation functions as limits of random variables more mysterious: an interpretation of non-bosonic fields, and in particular fermionic fields, as limit of discrete probabilistic objects is still accessible or not?

The aim of this short note is to understand that, at least in some cases, an interpretation of correlation functions of fermions as probabilistic objects is indeed still accessible. We are going to concentrate on the critical FK-Ising model, which converges in the scaling limit to the Ising CFT. This model possesses a natural and elegant geometrical representation that will allow to picture insertion of discrete fermions as {\it non-local} complex twists of topological events. More precisely, in the fully packed loop representation of the FK-Ising model, the correlation function of discrete fermionic observables counts the configurations in which insertion points are pairwise connected by loops, and it averages them with a complex factor depending on the winding of such loops. For instance, the FK-Ising two-point discrete fermionic observable will be given by
$$f(\z_{1},\z_{2})=\E_\text{FK}[{\bf1}_{\ga:\z_{1}\to\z_{2}}e^{-\frac i2 {\bf w}(\ga:\z_{1}\to\z_{2})}]$$ 

The interpretation of discrete fermions as topological objects that twist the probability measure can be originally dated back to the paper by Kadanoff and Ceva on defect lines \cite{kadanoff1971determination}, where they showed that for the Ising model, insertion of fermions corresponds to insertion of topological defect lines and the anti-commuting nature of the correlation functions is related to the winding of these defect lines. More recent developments with discrete complex analysis techniques have then led to a mathematically fulfilling connection between the discrete correlation of the critical Ising model to their continuum counterpart in the Ising CFT \cite{chelkak2015conformal,hongler2013energy}. More in general it has been understood that discrete holomorphic observables are a bridge between lattice models at criticality, CFT and Schramm Loewner Evolutions \cite{riva2006holomorphic,ikhlef2009discretely,alam2014integrability}.
In parallel, discrete fermionic observables for the critical FK-Ising model have been introduced by Smirnov \cite{smirnov2007towards,smirnov2010conformal,chelkak2012universality} as a fundamental tool to prove conformal invariance of the Ising model at criticality. 

Finally, it is worth pointing out that such a geometric interpretation of fermions does not rely on the discrete nature of the models, but it offers also insight of the nature of fermions at the continuum level, where Ising CFT fermion correlation functions are related to Schramm Loewner Evolution (SLE) martingales. In fact, {\it mutatis mutandis}, the picture of a fermionic correlation function as a complex twist of the expectation of the event of insertion points pairwise connected by loops still holds in the continuum where discrete paths are replaced by CLE loops. Indeed, just like in the Ising model correlations can be written as averages over geometrical configurations, either in terms of the FK-Ising model or the O$(1)$ model (low-temperature expansion); in the framework of the CFT/SLE correspondence, bCFT correlation functions are understood as averages over configurations sampled with SLE measures \cite{bauer2003conformal,bauer2003sle,bauer2005multiple,kytola2006conformal,hongler2013ising,dubedat2015sle}. In these terms, inserting a fermion $\psi(z)$ in bCFT correlation functions corresponds to gauge the SLE measures on paths that go through the insertion point $z$.

In the present note we are then going to extend Smirnov's definition of the observable to arbitrarily many insertion points; such an extension will be in a one-to-one correspondence with the Ising observable, as it appears in \cite{hongler2013conformal}. The structure of the paper is the following: in \secref{sec:fkmodel} we recall the critical FK-Ising model and its connection with the Ising model via the so-called Edwards-Sokal coupling; in \secref{sec:two-point} we introduce the winding phase and construct the two-point discrete fermionic observable for the FK-Ising model, and show its equivalence with the Ising observable as defined in \cite{honglerthesis}; in \secref{sec:many-point} and \secref{Discrete holomorphicity} we then extend such equivalence to the many-point case and observe that it possesses a pfaffian structure that should recall the reader of the free nature of the CFT Ising fermion. Finally, although a more complete and detailed analysis is deferred to a sequent note, in \secref{sec:beyond} we discuss on how to extend the construction for discrete fermions to its CLE counterpart.

\section*{Acknowledgments}
The author thanks Cl\'ement Hongler, Franck Gabriel and Sung Chul Park for helpful discussions; the author is supported by the ERC SG CONSTAMIS grant.

\section{FK-Ising model}\label{sec:fkmodel}
The family of Fortuin-Kasteleyn models, also known as random-cluster models, was introduced in the sixties as a unification of  edge-percolation, Ising, and Potts models. During the years, it has been extensively studied and we refer the reader to \cite{grimmett} for a rich overview of its properties and their proofs, as well as to \cite{smirnov2007towards,schramm2011conformally} for discussions on scaling limits. This section is dedicated to recalling the definition of the model, and to fixing the relevant notation that arise in the paper.

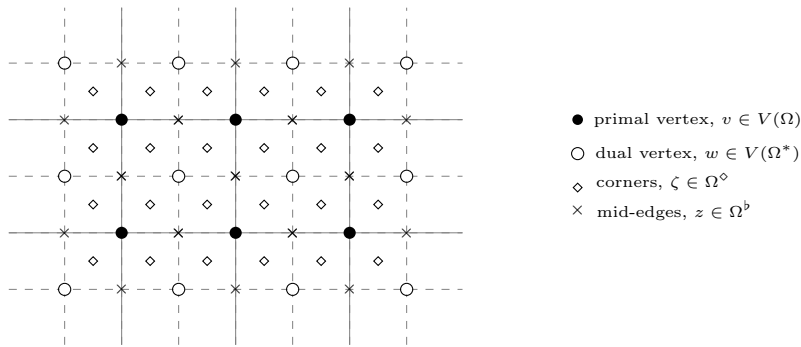
\begin{figure}[t]
\begin{tikzpicture}[scale=1.5]
\draw[step=1cm,gray,very thin] (-.99,-.99) grid (2.99,1.99);
\draw[step=.5cm,gray,very thin,dashed] (-.99,-.99) grid (2.99,1.99);
     \foreach \x in {0,...,2}{
        \foreach \y in {0,...,1}{
            \fill[black](\x,\y)circle(0.08/1.5);}
    }
         \foreach \x in {0,...,2}{
        \foreach \y in {0,...,1}{
            \node at (\x-.5,\y) {\tiny $\times$};
            \node at (\x+.5,\y) {\tiny $\times$};  
            \node at (\x,\y+.5) {\tiny $\times$};  
            \node at (\x,\y-.5) {\tiny $\times$};                
    }}
     \foreach \x in {-1,...,2}{
        \foreach \y in {0,...,2}{
            \fill[white](\x,\y)+(.5,-.5)circle(0.07/1.5);
                 \draw[black](\x,\y)+(.5,-.5)circle(0.08/1.5);}
        }
   \foreach \x in {0,...,2}{
        \foreach \y in {0,...,1}{
        \node[diamond,draw,xscale=.25,yscale=.25] at (\x+.25,\y+.25) {};  
      	\node[diamond,draw,xscale=.25,yscale=.25] at (\x+.25,\y-.25) {};        
       	\node[diamond,draw,xscale=.25,yscale=.25] at (\x-.25,\y+.25) {};      
	\node[diamond,draw,xscale=.25,yscale=.25] at (\x-.25,\y-.25) {};          
        }}
            \fill[black](4,1)circle(0.07/1.5) node {\qquad\qquad\qquad\qquad\quad \tiny primal vertex, $v\in V(\Om)$};
                        \fill[white](4,.7)circle(0.07/1.5);
                        \draw[black](4,.7)circle(0.08/1.5) node {\qquad\qquad\qquad\qquad\ \ \ \tiny dual vertex, $w\in V(\dual)$};
                        	\draw node[diamond,draw,xscale=.25,yscale=.25] at (4,.4) {};
	                        	\draw (4.75,.45) node {\tiny corners, $\z\in \corner$}; 
		                        	\node at (4,.2) {\tiny $\times$};
	                        	\draw (4.85,.2) node {\tiny mid-edges, $z\in \midpoint$}; 
\end{tikzpicture}
\caption{\label{fig:corners} A section of the lattice $\Om\subset\Z^{2}$ with its vertices $V(\Om)$ (black points) and its edges $E(\Om)$ (full lines), together with its dual lattice $\dual\simeq\Z^{2}$ with its vertices $V(\dual)$ (white points) and its edges $E(\dual)$. Diamonds $\diamond$ indicate corner points $\z\in\corner$, crosses $\times$ indicate mid-edges $z\in\midpoint$.}
\end{figure}

\subsec{Notation} For a sake of simplicity, throughout the whole paper we set our discussion on a finite region $\Om=(V(\Om),E(\Om))$ of the square lattice $\Z_{2}$, although the results can be straightforwardly extended to a wider class of graphs, e.g  isoradial graphs \cite{beffara2015,chelkak2011discrete}. Let us introduce the dual lattice $\dual$, the set of {\it corner points} $\corner$ consisting of the midpoints between two adjacent primal and dual vertices, and the set of {\it mid-edges} $z\in\midpoint$ consisting the midpoints between two adjacent primal vertices, as in \figref{fig:corners}. Each corner point $\z\in\corner$ is equivalently defined by a pair $(u,w)\in V(\Om)\times V(\dual)$ of adjacent primal and dual vertices, and as such it comes with a natural {\it orientation} $$o(\z)={w-u\over|w-u|} ,$$ note that $o(\z)\in\{e^{i\frac\pi4},-e^{i\frac\pi4},e^{-i\frac\pi4},-e^{-i\frac\pi4}\}$. We identify the corner $i$ and its pair of primal and dual vertices by writing $\z_{i}=(u_{i},w_{i})$.

\subsec{FK-Ising model} An FK configuration $\om\in\{0,1\}^{E(\Om)}$ consists of a collection of ``open'' and ``closed'' edges, respectively labeled with binary values $1$ and $0$.
The FK$_{2}$ model --also known as FK-Ising model, for the coupling that it possesses with the Ising model-- is a probability measure on subgraphs of $\Om$, defined, for all FK configurations $\om\in\{0,1\}^{E(\Om)}$ by
$$\rho_{p}(\om)={1\over \sZ_\text{FK}}{\left(p\over 1-p\right)}^{|\om|}2^{k(\om)}$$
where $|\om|$ is the number of open edges, $k(\om)$ is the number of cluster of primal vertices, $p\in[0,1)$ and $\sZ_\text{FK}$ is the partition function of the model, i.e. the normalization constant such that $\rho(\om)$ is a probability measure.

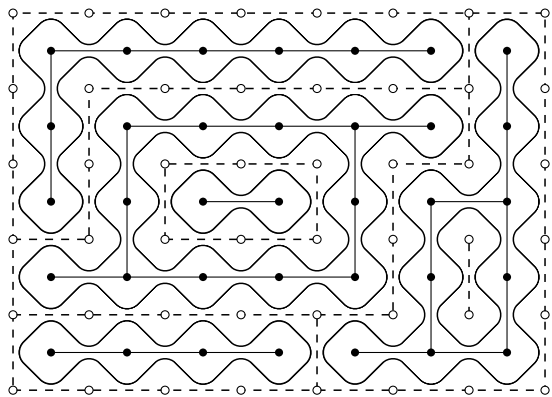
\begin{figure}[t]
\begin{tikzpicture}
    \foreach \x in {0,...,6}{
        \foreach \y in {0,...,4}{
            \draw[rounded corners=.2cm,line width=.02cm](\x,\y)+(.5,0)--+(0,.5)--+(-.5,0)--+(0,-.5)--cycle;
        }
    }
    \draw[draw dumbbells](0,1)--++(1,0)--++(0,2)--++(4,0)(1,1)--++(3,0)--++(0,2);
        \draw[draw dumbbells](2,2)--++(1,0);
            \draw[draw dumbbells](0,0)--++(3,0);
                \draw[draw dumbbells](0,2)--++(0,2)--++(5,0);
                            \draw[draw dumbbells](4,0)--++(2,0)--++(0,4)(5,0)--++(0,2)--++(1,0);
   \draw[line width=.02cm,dashed](-.5,-.5)--++(7,0)--++(0,5)--++(-7,0)--++(0,-5);
   \draw[line width=.02cm,dashed](-.5,.5)--++(5,0)--++(0,2)--++(1,0)--++(0,2)(-.5,1.5)--++(1,0)--++(0,2)--++(5,0)(3.5,-.5)--++(0,1);
   \draw[line width=.02cm,dashed](5.5,.5)--++(0,1);
      \draw[line width=.02cm,dashed](1.5,1.5)--++(2,0)--++(0,1)--++(-2,0)--++(0,-1);
    \foreach \x in {0,...,6}{
        \foreach \y in {0,...,4}{
            \fill[black](\x,\y)circle(0.08/1.5);
        }
    }
     \foreach \x in {0,...,7}{
        \foreach \y in {0,...,5}{
            \fill[white](\x,\y)+(-.5,-.5)circle(0.07/1.5);
                 \draw[black](\x,\y)+(-.5,-.5)circle(0.08/1.5);}
        }
\end{tikzpicture}
\caption{\label{fig:fk-config} Example of an FK configuration on a finite region of $\Z_{2}$ with five primal clusters, three dual clusters and seven loops separating them. In black vertices of $\Om$ and in white vertices in the dual lattice $\Om^{*}$. Open edges in the primal lattice are drawn with full stroke, while open edges in the dual lattice are drawn with dashes.}
\end{figure}

Any configuration $\om\in\{0,1\}^{E(\Om)}$ of edges is in a one-to-one correspondence with a configuration of edges in the dual lattice $\om^{*}\in\{0,1\}^{E(\Om^{*})}$: 
for any edge $e\in E(\Om)$ and its dual edge $e^{*}$, $$\om^{*}(e^{*})=1-\om(e) .$$ An useful representation of FK configurations is then obtained by separating clusters of primal edges and clusters of dual edges with {\it loops along corners}. In this representation, the probability measure can be rewritten as
$$\rho_{p}(\om)\propto t^{|\om|}\sqrt{2}^{\ell(\om)},$$
where $t={1\over \sqrt 2}{p\over 1-p}$ and $\ell(\om)$ is the number of loops around corners in the configuration $\om$. The FK-Ising model is critical at the value $t=1$, i.e. $p={\sqrt2 \over 1+\sqrt2}$ where the system becomes self-dual \cite{beffara2012self}.
In \figref{fig:fk-config} a typical configuration together with its loop representation is drawn.

In this paper we focus our attention to the FK model with free boundary conditions, i.e. with fully connected dual boundary edges, as in \figref{fig:fk-config}. This ensures that configurations consist of loops only (unlike the case of mixed boundary conditions --e.g. Dobrushin boundary conditions--, no open path is present). However, our construction of discrete fermionic observable can be easily extended in the case of different boundary conditions and we discuss about it in \ref{boundary conditions}.

If two vertices $u_{1},u_{2}\in V(\Om)$ are connected, i.e. belong to the same primal cluster, we will write $u_{1}\leftrightarrow u_{2}$; we use an analogous notation for connection of dual vertices and for corners.

\begin{lemma}\label{loop events}
Consider a FK configuration $\om$. Given two corners $\z_{1},\z_{2}\in\corner$, $\z_{1}=(u_{1},w_{1})$, $\z_{2}=(u_{2},w_{2})$ there exists a path along corners connecting $\z_{1}$ and $\z_{2}$ if and only if the primal vertices $u_{1}$, $u_{2}$ belong to the same primal cluster and the dual vertices $w_{1}$, $w_{2}$ belong to the same dual cluster. Equivalently,
\begin{equation}\label{}
\begin{split}
{\bf 1}_{\z_{1}\leftrightarrow\z_{2}}={\bf 1}_{u_{1}\leftrightarrow u_{2}}{\bf 1}_{w_{1}\leftrightarrow w_{2}}
\end{split}
\end{equation}
\end{lemma}
\proof 
We note that $u_{1}$ and $u_{2}$ do not belong to the same primal cluster, i.e. ${\bf 1}_{u_{1}\leftrightarrow u_{2}}=0$, if and only if there is a dual cluster $D$ separating them, i.e. without loss of generality $u_{1}$ is surrounded by dual edges of $D$ and $u_{2}$ is not. Thus, also the corner point $\z_{1}$ is surrounded by dual edges of $D$ and $\z_{2}$ is not. This implies that $\z_{1}$ and $\z_{2}$ are not connected, ${\bf 1}_{\z_{1}\leftrightarrow\z_{2}}=0$. By exchanging the role between primal and dual lattices, we have
$${\bf 1}_{\z_{1}\leftrightarrow\z_{2}}\leq{\bf 1}_{u_{1}\leftrightarrow u_{2}}{\bf 1}_{w_{1}\leftrightarrow w_{2}} .$$
If ${\bf 1}_{u_{1}\leftrightarrow u_{2}}{\bf 1}_{w_{1}\leftrightarrow w_{2}}=1$ then there are neither primal nor dual clusters separating $\z_{1}$ and $\z_{2}$, and thus ${\bf 1}_{\z_{1}\leftrightarrow\z_{2}}=1$.
\endproof

\subsection{Ising model and Edwards-Sokal coupling}
We recall the definition of the Ising model on vertices of $\Om$: to each vertex $x\in V(\Om)$ a binary spin $\s_{x}\in\{\pm1\}$ with a probability measure given by
$$\pi_{\beta}(\s)=\frac1\sZ_\text{Ising}e^{\beta\sum_{x\sim y}\s_{x}\s_{y}}$$
where the sum runs over neighboring sites, $\beta$ is a non-negative real parameter and $\sZ_\text{Ising}=\sum_{\s}e^{\beta\sum_{x\sim y}\s_{x}\s_{y}}$ is the partition function of the Ising model.

The Edwards-Sokal coupling \cite{edwards1988generalization} is a probability coupling of particular interest through which one can construct both the FK-Ising model and the Ising model on a common probability space. Precisely one consider configurations $(\s,\om)\in\{\pm1\}^{V(\Om)}\times\{0,1\}^{E(\Om)}$ and assign them a probability measure
\begin{equation}\label{}
\begin{split}
\mu(\s,\om)\propto\prod_{e\in E(\Om)}\left((1-p)\de_{\om(e),0}+p\de_{\om(e),1}\de_{e}(\s)\right)
\end{split}
\end{equation}
where $\de_{e}(\s)=\de_{\s_{x},\s_{y}}=\frac12(1+\s_{x}\s_{y})$ for $e=\bra x,y\ket\in E(\Om)$, and $p=1-e^{-2\beta}$.
The importance of this coupling relies on the fact of the following two aspects, for the proofs of which we refer the reader to \cite{grimmett},
\begin{itemize}
\item the marginal distributions of $\mu$ coincide with the FK-Ising and Ising measures:
$$\pi_{\beta}(\s)=\sum_{\om\in\{0,1\}^{E(\Om)}}\mu(\s,\om) ; \qquad \rho_{p}(\om)=\sum_{\s\in\{\pm1\}^{V(\Om)}}\mu(\s,\om) \ ;$$
\item Ising spin correlation corresponds to FK-Ising connection probabilities:
$$\bra\s_{x}\s_{y}\ket:=\sum_{\s\in\{\pm1\}^{V(\Om)}}\s_{x}\s_{y}\pi_{\beta}(\s)=\sum_{\om\in\{0,1\}^{E(\Om)}}{\bf 1}_{x\leftrightarrow y}(\om)\rho_{p}(\om)={\bb P}_{\rho}[x\longleftrightarrow y] .$$
\end{itemize}
For $\om\in\{0,1\}^{E(\Om)}$ the conditional measure $\mu(\, \cdot\, |\om)$ on $\{\pm1\}^{V(\Om)}$ can be obtained by assigning, with equal probability, random $\pm1$ spins on entire clusters of $\om$. These spins are constant on given clusters, and independent between different clusters. For $\s\in\{\pm1\}^{V(\Om)}$ the conditional measure $\mu(\,\cdot\,|\s)$ on $\Om$ is obtained by flipping an independent biased coin for each edge $e=\bra x,y\ket$ between two adjacent vertices $x,y\in V(\Om)$ with same spin $\s_{x}=\s_{y}$, and assigning $\om(e)=1$ with probability $p$ and $0$ otherwise.

\subsection{Disorder lines} The main tool in the study of discrete Ising fermions is given by 
the introduction of disorder lines; these correspond to the results of the insertion of discrete disorder operators \cite{kadanoff1971determination}. An {\it a posteriori} intuition for introducing such an object is that in the Ising CFT the Ising fermion $\psi$ can be seen as the product of a pair of spin and disorder field; similarly, in the Ising model, the discrete fermionic will be given on a lattice algebra by the product of a spin and a disorder operator \cite{mussardo2010statistical,francesco2012conformal,hongler2013conformal}.

\begin{definition}[Disorder line]
By a disorder line between two dual vertices $p, q \in \dual$ we mean a simple path $\la$ along dual edges with end-points $p,q$. For an Ising configuration on primal vertices $(\s_{x})_{x\in\Om}$ define the {\it disorder energy} $E_{\la}[\s]$ as $$E_{\la}[\s]=\sum_{x\sim y:\bra xy\ket^{*} \in \la}\s_{x}\s_{y}\ ,$$ where the sum is over all primal edges $\bra xy\ket$ orthogonal to dual edges of $\la$. For a disorder line between $p$ and $q$ define the {\it disorder pair} $(\mu_{p}\mu_{q})_{\la}$ as $\exp(-2\beta E_{\la}[\s])$.
\end{definition}

\begin{figure}[t]
\begin{subfigure}[b]{0.49\textwidth}
\centering
\begin{tikzpicture}
    \foreach \x in {0,...,6}{
        \foreach \y in {0,...,4}{
            \draw[rounded corners=.2cm,line width=.02cm](\x,\y)+(.5,0)--+(0,.5)--+(-.5,0)--+(0,-.5)--cycle;
        }
    }
    \draw[draw dumbbells](0,1)--++(1,0)--++(0,2)--++(4,0)(1,1)--++(3,0)--++(0,2);
        \draw[draw dumbbells](2,2)--++(1,0);
            \draw[draw dumbbells](0,0)--++(3,0);
                \draw[draw dumbbells](0,2)--++(0,2)--++(5,0);
                            \draw[draw dumbbells](4,0)--++(2,0)--++(0,4)(5,0)--++(0,2)--++(1,0);
   \draw[line width=.02cm,dashed](-.5,-.5)--++(7,0)--++(0,5)--++(-7,0)--++(0,-5);
   \draw[line width=.02cm,dashed](-.5,.5)--++(5,0)--++(0,2)--++(1,0)--++(0,2)(-.5,1.5)--++(1,0)--++(0,2)--++(5,0)(3.5,-.5)--++(0,1);
   \draw[line width=.02cm,dashed](5.5,.5)--++(0,1);
      \draw[line width=.02cm,dashed](1.5,1.5)--++(2,0)--++(0,1)--++(-2,0)--++(0,-1);
    \foreach \x in {0,...,6}{
        \foreach \y in {0,...,4}{
 \node[circle,draw,xscale=.35,yscale=.35,fill=orange] at(\x,\y) {{$\bf +$}};
        }
    }
\node[circle,draw,xscale=.35,yscale=.35,fill=darkgreen] at(0,0) {{$\bf -$}};    
\node[circle,draw,xscale=.35,yscale=.35,fill=darkgreen] at(1,0) { {$\bf -$}};      
\node[circle,draw,xscale=.35,yscale=.35,fill=darkgreen] at(2,0) { {$\bf -$}};   
\node[circle,draw,xscale=.35,yscale=.35,fill=darkgreen] at(3,0) { {$\bf -$}};   
\node[circle,draw,xscale=.35,yscale=.35,fill=darkgreen] at(2,2) { {$\bf -$}};   
\node[circle,draw,xscale=.35,yscale=.35,fill=darkgreen] at(3,2) { {$\bf -$}};   
\node[circle,draw,xscale=.35,yscale=.35,fill=darkgreen] at(0,2) { {$\bf -$}};   
\node[circle,draw,xscale=.35,yscale=.35,fill=darkgreen] at(0,3) { {$\bf -$}};   
\node[circle,draw,xscale=.35,yscale=.35,fill=darkgreen] at(0,4) { {$\bf -$}};   
\node[circle,draw,xscale=.35,yscale=.35,fill=darkgreen] at(1,4) { {$\bf -$}};   
\node[circle,draw,xscale=.35,yscale=.35,fill=darkgreen] at(2,4) { {$\bf -$}};   
\node[circle,draw,xscale=.35,yscale=.35,fill=darkgreen] at(3,4) { {$\bf -$}};   
\node[circle,draw,xscale=.35,yscale=.35,fill=darkgreen] at(4,4) { {$\bf -$}};   
\node[circle,draw,xscale=.35,yscale=.35,fill=darkgreen] at(5,4) { {$\bf -$}};   
     \foreach \x in {0,...,7}{
        \foreach \y in {0,...,5}{
            \fill[white](\x,\y)+(-.5,-.5)circle(0.07/1.5);
                 \draw[black](\x,\y)+(-.5,-.5)circle(0.08/1.5);}
        }
\end{tikzpicture}
\end{subfigure}
\begin{subfigure}[b]{0.49\textwidth}
\centering
\begin{tikzpicture}
    \foreach \x in {0,...,6}{
        \foreach \y in {0,...,4}{
            \draw[rounded corners=.2cm,line width=.02cm](\x,\y)+(.5,0)--+(0,.5)--+(-.5,0)--+(0,-.5)--cycle;
        }
    }
    \draw[draw dumbbells](0,1)--++(1,0)--++(0,2)--++(4,0)(1,1)--++(3,0)--++(0,2);
        \draw[draw dumbbells](2,2)--++(1,0);
            \draw[draw dumbbells](0,0)--++(3,0);
                \draw[draw dumbbells](0,2)--++(0,2)--++(5,0);
                            \draw[draw dumbbells](4,0)--++(2,0)--++(0,4)(5,0)--++(0,2)--++(1,0);
   \draw[line width=.02cm,dashed](-.5,-.5)--++(7,0)--++(0,5)--++(-7,0)--++(0,-5);
   \draw[line width=.02cm,dashed](-.5,.5)--++(5,0)--++(0,2)--++(1,0)--++(0,2)(-.5,1.5)--++(1,0)--++(0,2)--++(5,0)(3.5,-.5)--++(0,1);
   \draw[line width=.02cm,dashed](5.5,.5)--++(0,1);
      \draw[line width=.02cm,dashed](1.5,1.5)--++(2,0)--++(0,1)--++(-2,0)--++(0,-1);
    \foreach \x in {0,...,6}{
        \foreach \y in {0,...,4}{
 \node[circle,draw,xscale=.35,yscale=.35,fill=orange] at(\x,\y) {{$\bf +$}};
        }
    }
\node[circle,draw,xscale=.35,yscale=.35,fill=darkgreen] at(0,0) {{$\bf -$}};    
\node[circle,draw,xscale=.35,yscale=.35,fill=darkgreen] at(1,0) { {$\bf -$}};      
\node[circle,draw,xscale=.35,yscale=.35,fill=darkgreen] at(2,0) { {$\bf -$}};   
\node[circle,draw,xscale=.35,yscale=.35,fill=darkgreen] at(3,0) { {$\bf -$}};   
\node[circle,draw,xscale=.35,yscale=.35,fill=darkgreen] at(2,2) { {$\bf -$}};   
\node[circle,draw,xscale=.35,yscale=.35,fill=darkgreen] at(3,2) { {$\bf -$}};   
\node[circle,draw,xscale=.35,yscale=.35,fill=darkgreen] at(0,2) { {$\bf -$}};   
\node[circle,draw,xscale=.35,yscale=.35,fill=darkgreen] at(0,3) { {$\bf -$}};   
\node[circle,draw,xscale=.35,yscale=.35,fill=darkgreen] at(0,4) { {$\bf -$}};   
\node[circle,draw,xscale=.35,yscale=.35,fill=darkgreen] at(1,4) { {$\bf -$}};   
\node[circle,draw,xscale=.35,yscale=.35,fill=darkgreen] at(2,4) { {$\bf -$}};   
\node[circle,draw,xscale=.35,yscale=.35,fill=darkgreen] at(3,4) { {$\bf -$}};   
\node[circle,draw,xscale=.35,yscale=.35,fill=darkgreen] at(4,4) { {$\bf -$}};   
\node[circle,draw,xscale=.35,yscale=.35,fill=darkgreen] at(5,4) { {$\bf -$}};   
     \foreach \x in {0,...,7}{
        \foreach \y in {0,...,5}{
            \fill[white](\x,\y)+(-.5,-.5)circle(0.07/1.5);
                 \draw[black](\x,\y)+(-.5,-.5)circle(0.08/1.5);}
        }
        \draw[line width=.03cm,decorate,decoration={snake,amplitude=.03cm,segment length=2mm,post length=0},cambridge](0.5,1.5)--++(0,-2)--++(1,0)--++(0,4)--++(1,0)--++(0,-3)--++(2,0)--++(0,1);
\node[circle,draw,xscale=.35,yscale=.35,fill=darkgreen] at(1,1) { {$\bf -$}};      
\node[circle,draw,xscale=.35,yscale=.35,fill=darkgreen] at(1,2) { {$\bf -$}};      
\node[circle,draw,xscale=.35,yscale=.35,fill=darkgreen] at(1,3) { {$\bf -$}};      
\node[circle,draw,xscale=.35,yscale=.35,fill=darkgreen] at(3,3) { {$\bf -$}};      
\node[circle,draw,xscale=.35,yscale=.35,fill=darkgreen] at(4,3) { {$\bf -$}};      
\node[circle,draw,xscale=.35,yscale=.35,fill=darkgreen] at(5,3) { {$\bf -$}};      
\node[circle,draw,xscale=.35,yscale=.35,fill=darkgreen] at(4,2) { {$\bf -$}};      
\node[circle,draw,xscale=.35,yscale=.35,fill=darkgreen] at(4,1) { {$\bf -$}};      
\node[circle,draw,xscale=.35,yscale=.35,fill=darkgreen] at(3,1) { {$\bf -$}};      
\node[circle,draw,xscale=.35,yscale=.35,fill=orange] at(3,2) { {$\bf +$}};      
\node[circle,draw,xscale=.35,yscale=.35,fill=orange] at(1,0) { {$\bf +$}};      
\end{tikzpicture}
\end{subfigure}
\caption{\label{fig:edwards-sokal} Left: an Ising configuration sampled from the FK configuration of \figref{fig:fk-config}: for each of the five clusters, independently, one assigns to the whole cluster $\pm1$ spins according to the outcome of a fair coin. Right: the same FK configuration in the presence of a disorder line (in red): spins are assign as before by independent tossing of a fair coin, but regions of a cluster separated by the disorder line have opposite spin.}
\end{figure}
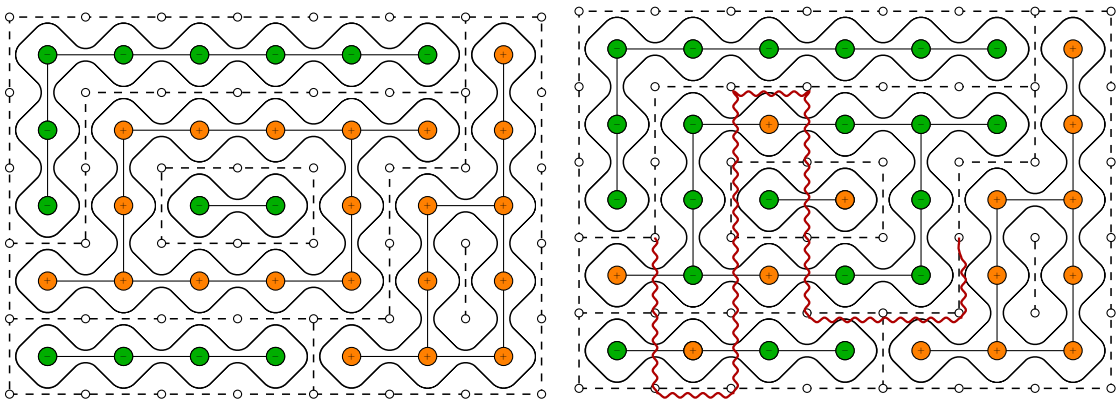

In terms of Ising correlation functions, the introduction of a disorder pair $(\mu_{p}\mu_{q})_{\la}$, is equivalent to the effect of changing along primal edges crossing $\la$ the Ising model from ferromagnetic, with parameter $\beta$, to antiferromagnetic, with parameter $-\beta$:
 typical Ising configuration with a disorder line $\la$ would tend to favor alignment of spins away from the disorder line $\la$, but opposite alignment of spins along $\la$.

We want then to modify the Edwards-Sokal coupling so to take into account the presence of a disorder line between $w_{1}$ and $w_{2}$. Precisely, we  modify the coupling so that if a primal cluster is separated by $\la$ in different regions, spins would be equal throughout the region, but opposite between  two adjacent regions, see \figref{fig:edwards-sokal}. To take the following property of disorder line into account, one modifies the above coupling with the {\it Edwards-Sokal coupling with disorder line $\la$} as
$$\mu_{\la}(\s,\om)\propto\prod_{e\in E(\Om)\setminus\la}\left((1-p)\de_{\om(e),0}+p\de_{\om(e),1}\de_{e}(\s)\right)\prod_{e\in \la}\left((1-p)\de_{\om(e),0}+p\de_{\om(e),1}(1-\de_{e}(\s))\right)$$
Any configuration $\om$ in which $w_{1}$ and $w_{2}$ are not connected, i.e. in which there is a primal cluster surrounding $w_{1}$ but not $w_{2}$, or vice versa, 
is such that $\mu_{\la}(\s,\om)=0$ for any spin configuration $\s$. In fact, $\la$ would necessarily cut through the primal cluster, and any spin configuration will result to $\de_{e}(\s)=0$ away from $\la$ or $\de_{e}(\s)=1$ across $\la$. So one has
$$\sum_{\om\in\{0,1\}^{E(\Om)}}\mu_{\la}(\s,\om)=e^{-2\beta E_{\la}[\s]}\pi_{\beta}(\s) \ , 
 \qquad\sum_{\s\in\{\pm1\}^{V(\Om)}}\mu_{\la}(\s,\om)= \rho_{p}(\om){\bf 1}_{w_{1}\to w_{2}} \ .$$

\section{Two-point fermionic observable}\label{sec:two-point}
In this section we introduce the discrete fermionic observable with two insertion points. As in \cite{smirnov2010conformal}, the observable would be defined on corner points $\z\in\corner$. The fermionic observable is an average, with respect to the FK measure, of complex indicator of (non-local) connection events. The non-locality nature of the observable comes from the introduction of the winding phase: a measure of how much a path moving from one corner to another goes around its end corner, or equivalently how much the tangent vector field along the curve turns. We define the winding phase with the convention that 
{\it loops are walked by keeping primal clusters to their left}.

\subsection{Winding}
\begin{definition}[Winding phase]
Suppose $\z_{1},\z_{2}\in\corner$ are distinct corner points connected by a path $\ga$, the winding ${\bf w}(\ga: \z_{1}\to \z_{2})$ of $\ga$ from $\z_{1}$ to $\z_{2}$ is defined as the total angle that the path $\ga$ (walked by keeping primal clusters to its left) takes to go from $\z_{1}$ to $\z_{2}$
\begin{equation}\label{}
\begin{split}
{\bf w}(\ga: \z_{1}\to \z_{2})= \frac\pi2(n_\text{right}-n_\text{left}) .
\end{split}
\end{equation}
The {\it winding phase} $\phi\in\{\pm1\}$ of $\ga$ from $\z_{1}$ to $\z_{2}$ as
\begin{equation}\label{}
\begin{split}
\phi(\ga,\z_{1},\z_{2})=\sqrt{\frac{o(\z_{1})}{o(\z_{2})}}\exp\(-\frac i2 {\bf w}(\ga: \z_{1}\to \z_{2})\) .
\end{split}
\end{equation}
\end{definition}
By definition, an FK loops keeps the boundary of a primal cluster on its left, thus the winding ${\bf w}(\ga:\z_{1}\to\z_{2})$ can be equivalently computed either moving along the path $\ga$ or along the boundary (primal edges) of the primal cluster on its left from $u_{1}$ to $u_{2}$ or along the boundary (dual edges) of the dual cluster on its right from $w_{1}$ to $w_{2}$.

\begin{figure}[t]
%
%
%
\def\svgwidth{.4\textwidth}
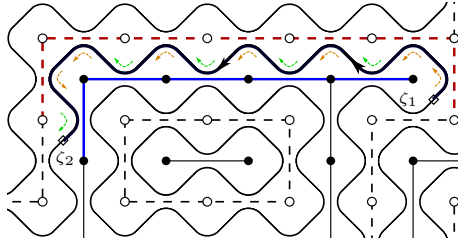
\caption{\label{fig:winding-config} In bold black, the path $\ga$ from $\z_{1}$ to $\z_{2}$ on a FK configuration. The winding of $\ga$ moves five times on the right (green turnings) and seven times on the left (orange turnings), so that its total winding is $\frac\pi2(5-7)=-\pi$.}
\end{figure}
In the next result we focus on the order on which the entries appear when walking along a loop, in this case the insertion points will be denoted with $\z^{i}$ to indicate that it is the $i$-th insertion point visited on the loop exploration.

\begin{proposition}\label{prop:winding-properties}
For a loop $\ga$, the winding phase possess the following properties:
\begin{enumerate}
\item $\phi(\ga,\z_{1},\z_{2})$ is an antisymmetric functions in the variables $\z_{1},\z_{2}\in\corner$, i.e.
$$\phi(\ga,\z_{1},\z_{2})=-\phi(\ga,\z_{2},\z_{1})\ ;$$
\item for any triple of ordered distinct corner points $\z^{1},\z^{2},\z^{3}\in\corner$ laying on the same loop $\ga$, one has
$$\phi(\ga,\z^{1},\z^{2})\phi(\ga\z^{2},\z^{3})=\phi(\ga,\z^{1},\z^{3})\ .$$
\end{enumerate}
\end{proposition}
\proof{\white .} \\
\begin{enumerate}
\item The cluster configuration determines a loop $\ga_{\z_{1}\to\z_{2}}\cup\tilde\ga_{\z_{2}\to\z_{1}}$ that runs keeping the primal cluster on its left and that goes through the corners $\z_{1}$ and $\z_{2}$.
Since $${\bf w}(\ga:\z_{1}\to\z_{2})+{\bf w}(\tilde\ga:\z_{2}\to\z_{1})=2\pi ,$$ it follows that 
$$\frac{\phi(\ga,\z_{1},\z_{2})}{\phi(\tilde\ga,\z_{2},\z_{1})}=-\frac{o(\z_{1})}{o(\z_{2})}\exp\left(- i {\bf w}(\ga: \z_{1}\to \z_{2})\right) .$$
The orientation of $\z_{1}$ and $\z_{2}$ determines the value of the winding ${\bf w}(\ga:\z_{1}\to\z_{2})$: $\bf w$ is equal, modulo $2\pi$, to the difference of the phase of the orientations; thus
$$\exp\left(- i {\bf w}(\ga: \z_{1}\to \z_{2})\right)=\frac{o(\z_{2})}{o(\z_{1})} .$$
\item One has
$$\phi(\ga,\z^{1},\z^{2})\phi(\ga,\z^{2},\z^{3})=\sqrt{\frac{o(\z^{1})}{o(\z^{2})}}\sqrt{\frac{o(\z^{2})}{o(\z^{3})}}\exp\(-\frac i2 {\bf w}(\ga: \z^{1}\to \z^{2})\)\exp\(-\frac i2 {\bf w}(\ga: \z^{2}\to \z^{3})\)$$
and then the thesis follows from the definition of the orientation $o(\z)$ and of the winding ${\bf w}$.
\end{enumerate}
\endproof

The winding phase $\phi$ is the non-local term that allows to have an antisymmetric observable: for two distinct corner points this is defined as the average of the winding phase over all possible path connecting the two corner points. 

\begin{definition}[Two-point discrete fermionic observable]
Let $\z_{1},\z_{2}$ two distinct corner points, the two-point discrete fermionic observable is defined as
\begin{equation*}\label{}
\begin{split}
f(\z_{1},\z_{2}):&=\E[{\bf 1}_{\ga:\z_{1}\rightarrow\z_{2}}\phi(\ga,\z_{1},\z_{2})] \\
&=\sum_{\om}\rho_{p}(\om){\bf 1}_{\ga(\om):\z_{1}\rightarrow\z_{2}}\phi(\ga(\om),\z_{1},\z_{2}) ;
\end{split}
\end{equation*}
\end{definition}
%

It is important to notice that the antisymmetry of the real function is a consequence of the factor $\frac12$ in front of the winding $\bf w$, named {\it spin} in the literature. Any value different from a semi-integer it would not give rise to antisymmetric function: if the {\it spin} were integer one would get a discrete bosonic observable, i.e. expectation of random variables; if the spin is instead in $\Q\setminus \frac12\Z$ then one has more complicated observables, named {\it parafermionic} observables \cite{smirnov2007towards}. The discrete fermionic observable for the FK model was originally introduced by Smirnov on a complexified version 
$$F(\z_{1},\z_{2}):=\sqrt{i\over o(\z_{2})}\E[{\bf 1}_{\ga:\z_{1}\rightarrow\z_{2}}\phi(\ga,\z_{1},\z_{2})] ;$$
and for the case of Dobrushin boundary conditions, where the two boundary arcs between two pivotal boundary corner points $a$ and $b$ are fully connected with primal edges and with dual edges. Beyond loops, such a setting creates a path from $a$ to $b$, and the function $F(\z,a)$ is a martingale with respect to the filtration induced by the exploration of the path.

\subsection{Ising model two-point discrete fermionic observable}

As anticipated, the discrete fermionic observable for the Ising model has already been defined as a discrete holomorphic function that converges in the scaling limit to the fermion of the Ising CFT in the sense of correlation functions. We now recall the definition of  the Ising discrete fermionic observable  as in \cite{honglerthesis,hongler2013energy,hongler2013conformal,gheissari2019ising}.

\begin{definition}
Let $\z_{1}$ be a corner between $u_{1}\in V(\Om)$ and $w_{1}\in V(\dual)$, and let $\z_{2}$ be a corner between $u_{2}\in V(\Om)$ and $w_{2}\in V(\dual)$. Define a {\it corner defect line $\la$ with corner-ends} $\z_{1},\z_{2}\in\corner$, $\la:\z_{1}\to\z_{2}$, as the concatenation $[\z_{1}w_{1}]+\ga+[w_{2}\z_{2}]$ of a disorder line $\ga$ with endpoints $w_{1},w_{2}$ with the two {\it corner segments} $[\z_{1}w_{1}]$ and $[w_{2}\z_{2}]$. $u_{1},u_{2}$ are called the {\it spin-ends} and $w_{1},w_{2}$ the {\it disorder-ends}. Denote by ${\bf W}(\la:\z_{1}\to\z_{2})$ the {\it total turning} of $\la$ (also known as {\it winding}) when going from $\z_{1}$ to $\z_{2}$.
\end{definition}

Although both of them are functions defined on corners, the winding ${\bf w}(\ga:\z_{1}\to\z_{2})$ in the FK model  counts the winding of the FK loop connecting $\z_{1}$ to $\z_{2}$, while the winding ${\bf W}(\la: \z_{1}\to\z_{2})$ in the Ising model counts the winding of the disorder line, i.e. a line living on dual edges.

Windings of different defect lines having same corner-ends can be compared by studying their symmetric difference, and in particular its rotation number. For a closed curve, piecewise regular, parametrized curve $\alpha:[0,1]\to\C$, with vertices $\alpha(t_{i})$ and external angles $\theta_{i}$, $i=1,\dots,k$, let $\phi:[0,1]\setminus\{t_{i}\}_{i=1:k}\to\S^{1}$ be given by $\phi(t)={\alpha'(t)\over|\alpha'(t)|}$. Then the rotation number $\sR(\alpha)\in\Z$ is defined as $$2\pi\sR(\alpha):=\sum_{i=1}^{k}(\phi(t_{i+1})-\phi(t_{i}))+\sum_{i=1}^{k}\theta_{i} .$$ Intuitively, the rotation number measures the complete turns given by the tangent vector field along the curve \cite{do2016differential}.

\begin{lemma}\label{winding-interior}
Let $\z_{1},\z_{2}\in\corner$ be two corner points, and let $\la,\tilde\la$ be two corner defect line with corner-ends $\z_{1},\z_{2}\in\corner$. Let $\la\oplus\tilde\la$ denote the collection of loops made of the symmetrtic difference of $\la$ and $\tilde\la$. Then
$$e^{- \frac i2 {\bf W}(\la:\z_{1}\to\z_{2})}=(-1)^{\sN}e^{- \frac i2 {\bf W}(\tilde\la:\z_{1}\to\z_{2})}$$
where $\sN$ is the number of self-intersections of $\la\oplus\tilde\la$.
\end{lemma}
\proof
One has that $${\bf W}(\la:\z_{1}\to\z_{2})-{\bf W}(\tilde\la:\z_{1}\to\z_{2})={\bf W}(\la:\z_{1}\to\z_{2})+{\bf W}(\tilde\la:\z_{2}\to\z_{1})=\sR(\la\oplus\tilde\la)\pm2\pi\ ,$$ where the $\pm1$ sign depend on the orientation of the loop. And it is a straightforward consequence of Whitney's formula \cite{arnold1994topological,whitney1937regular} that the rotation number of a curve is an odd multiple of $2\pi$ if and only if the number of its self-intersections is even.
\endproof

\begin{definition}
Let $\la$ be a corner defect line with corner-ends $\z_{1},\z_{2}$, spin-ends $u_{1},u_{2}$ and disorder-ends $w_{1},w_{2}$. We define the {\it real} fermion pair $(\psi(\z_{1})\psi(\z_{2}))_{\la}$ as
$$(\psi(\z_{1})\psi(\z_{2}))_{\la}=i \sqrt{o(\z_{2})\over o(\z_{1})}e^{-\frac i2 {\bf W}(\la:\z_{1}\to \z_{2})}(\mu_{w_{1}}\mu_{w_{2}})_{\ga}\s_{u_{1}}\s_{u_{2}}=i \sqrt{o(\z_{2})\over o(\z_{1})}e^{-\frac i2 {\bf W}(\la:\z_{1}\to \z_{2})}e^{-2\beta E_{\ga}[\s]}\s_{u_{1}}\s_{u_{2}}$$
\end{definition}

Although we use the same notation $(\psi(\z_{1})\psi(\z_{2}))_{\la}$ as in \cite{hongler2013conformal}, our definition of the real fermion pair differs from their definition of (non-real) fermion pair by a factor of $-i \sqrt{o(\z_{1})}\sqrt{o(\z_{2})}$.

\begin{lemma}
Let $\la$ be a corner defect line with corner-ends $\z_{1},\z_{2}$. Then we have 
$$(\psi(\z_{1})\psi(\z_{2}))_{\la}+(\psi(\z_{2})\psi(\z_{1}))_{\la}=0$$
\end{lemma}
\proof Ref. sec. 3 of \cite{hongler2013conformal}.
\endproof

\begin{lemma}
Let $\la,\tilde \la$ be two corner defect lines sharing the same corner-ends $\z_{1},\z_{2}$. 
Then $$\bra (\psi(\z_{1})\psi(\z_{2}))_{\la} \ket =\bra (\psi(\z_{1})\psi(\z_{2}))_{\tilde\la} \ket . $$
\end{lemma}
\proof Ref. sec. 3 of \cite{hongler2013conformal}.
\endproof

\begin{figure}[t]
\begin{tikzpicture}
   \draw[step=1cm,gray,very thin,xshift=-0.5cm,yshift=-0.5cm] (0,0) grid (7,5);
   \draw[line width=.02 cm](-.5,.5)--++(1,0)--++(0,0.9)--++(0.1,0.1)--++(0.9,0)--++(0,1)--++(-1,0)--++(0,-0.9)--++(-0.1,-0.1)--++(-0.9,0)--++(0,-1);
   \draw[line width=.02 cm,darkblue](2.75,3.75)--++(-.25,-.25)--++(0,-1)--++(1,0)--++(0,-1)--++(0.9,0)--++(0.1,-0.1)--++(0,-1.9)--++(-2,0)--++(0,1)--++(1,0)--++(0.25,0.25);
      \draw[line width=.02 cm](3.5,3.5)--++(1,0)--++(0,-1.9)--++(0.1,-0.1)--++(1.9,0)--++(0,2)--++(-1,0)--++(0,1)--++(-2,0)--++(0,-1);
      \draw (2.75,3.9) node {\tiny $\z_{1}$};
            \draw (3.75,.9) node {\tiny $\z_{2}$};
\end{tikzpicture}
\caption{\label{fig:loop-config} A configuration $\ga\in\sC(\z_{1},\z_{2})$ of simple loops (in black), and a corner defect line $\pi(\ga)$ from $\z_{1}$ to $\z_{2}$ (in blue). The corner defect line $\pi(\ga)$ has winding ${\bf W}(\pi(\ga):\z_{1}\to\z_{2})=\pi$.}
\end{figure}
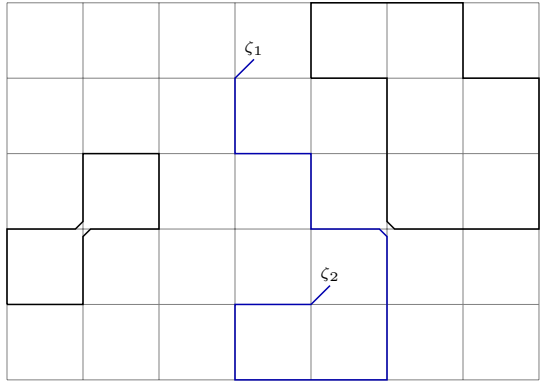

\begin{definition}[Ising two-point fermionic observable]
Consider two corner points $\z_{1},\z_{2}\in\corner$, 
then
\begin{equation}\label{}
\begin{split}
f_{\text{Ising}}(\z_{1},\z_{2})&=\bra (\psi(\z_{1})\psi(\z_{2}))_{\la} \ket = i \sqrt{o(\z_{2})\over o(\z_{1})}\frac1\sZ\sum_{\sC(\z_{1},\z_{2})}e^{-2\beta |\la|}e^{-\frac i2 {\bf W}(\la: \z_{1}\to \z_{2})}
\end{split}
\end{equation}
where $\sZ=\sum_{\ell\in\sC}e^{-2\beta|\ell|}=e^{-\beta|E|}\sum_{\s}e^{\beta\sum_{x\sim y}\s_{x}\s_{y}}=e^{-\beta|E|}\sZ_\text{Ising}$, and $\sC$ is the set of configuration consisting of dual loops, $\sC(\z_{1},\z_{2})$ is the set of configurations consisting of collections of dual loops and a corner defect line $\la$ from $\z_{1}$ and $\z_{2}$ (ref. \figref{fig:loop-config}); $|\la|$ is the sum of the lengths of the dual path connecting $\z_{1}$ to $\z_{2}$ and the length of all the loops in the configuration.
\end{definition}

\subsection{Equivalence between FK-Ising and Ising two-point observables}
Although, a priori different, the two-point observables defined for the critical FK-Ising model and for the critical Ising model coincide. In order to prove this statement, the general idea is on one side to relate the winding of the FK-Ising loop ${\bf w}(\ga:\z_{1}\to\z_{2})$ and the winding of the disorder line ${\bf W}(\la:\z_{1}\to\z_{2})$ and on the other side to relate the event that there exists a path $\ga:\z_{1}\to\z_{2}$ to the presence of the disorder line $\la$.

We are now in the position of proving the main theorem of the section:
\begin{theorem}\label{theorem:2pts-equivalence}
Let $\z_{1},\z_{2}\in\corner$ two distinct corner points. Then the discrete fermionic observables for the FK-Ising model and for the Ising model defined on $\z_{1}$ and $\z_{2}$ coincide.
\emph{$$f_\text{FK}(\z_{1},\z_{2})=f_\text{Ising}(\z_{1},\z_{2})$$}
\end{theorem}
\proof
We have already seen in Proposition \ref{loop events} that
$${\bf 1}_{\z_{1}\leftrightarrow\z_{2}}={\bf 1}_{u_{1}\leftrightarrow u_{2}}{\bf 1}_{w_{1}\leftrightarrow w_{2}}\ .$$
Thus one can fix an arbitrary corner defect line $\la$ from $\z_{1}$ to $\z_{2}$ and rewrite the FK weight $\rho_{p}(\om){\bf 1}_{w_{1}\leftrightarrow w_{2}}$ as the marginal measure of the Edwards-Sokal coupling measure $\mu_{\la}(\s,\om)$ with defect line $\la$.
Furthermore, the winding of the path $\ga:\z_{1}\to\z_{2}$ can be reformulated in terms of the winding of the path $\ga^{\text w}$ along the boundary of the dual cluster, and one has
$${\bf w}(\ga:\z_{1}\to\z_{2})={\bf W}(\ga^{\text w}: \z_{1}\to \z_{2})-\pi \ .$$

When a defect line $\la$ crosses $\ga^{\text w}$ it crosses also the boundary of the primal cluster, thus the number of self-crossings $\sN$ of $\la\oplus\ga^{\text w}$ corresponds to the number of forced spin-flippings in the primal cluster of $u_{1}$ and $u_{2}$; equivalently, $(-1)^{\sN}=\s_{u_{1}}\s_{u_{2}}$. Thanks to lemma \ref{winding-interior} and the Edwards-Sokal coupling with a defect line $\la$, one has that
$${\bf 1}_{u_{1}\to u_{2}}e^{-\frac i2{\bf W}(\ga^{\text w}: \z_{1}\to \z_{2})}=\s_{u_{1}}\s_{u_{2}}e^{-\frac i2{\bf W}(\la: \z_{1}\to \z_{2})} \ .$$
Again, by Lemma \ref{winding-interior}, given a simple path $\la$ and a collection of loops $\eta$ one has that the phase of the path in the symmetric difference $\la\oplus\eta$ is equal up to a sign to the phase of the path of $\la$:
$$e^{- \frac i2 {\bf W}(\la\oplus\eta:\z_{1}\to\z_{2})}=(-1)^{\sN}e^{- \frac i2 {\bf W}(\la:\z_{1}\to\z_{2})} .$$

In the low-temperature expansion for an Ising configuration of spins $\s$, configurations of dual loops $\eta$ are obtained as collection of dual edges $e\in E(\dual)$ across which the spins change signs, i.e. $\de_{e}(\eta)=\frac12\left(1-\s_{x}\s_{y}\right)$, where $x,y\in V(\Om)$ are vertices of the primal edge orthogonal to $e$.
The energy of the configuration $\s$ can then be written as
$$E[\s]=\sum_{x\sim y}\s_{x}\s_{y}=\sum_{e\in E(\dual)}(1-2\de_{e}(\eta))=|E|-2|\eta|$$
Furthermore,
$$\de_{e}(\la\oplus\eta)=\de_{e}(\la)+\de_{e}(\eta)-2\de_{e}(\la)\de_{e}(\eta) ,$$
and thus one has
\begin{equation*}\label{}
\begin{split}
-2 E_{\la}[\s]+ E[\s]&=-2\sum_{e\in E(\dual)}\de_{e}(\la)(1-2\de_{e}(\eta))+\sum_{e\in E(\dual)}\left(1-2\de_{e}(\eta)\right)
\\&=\sum_{e\in E(\dual)}\left(4\de_{e}(\la)\de_{e}(\eta)-2\de_{e}(\la)-2\de_{e}(\eta)+1\right)
\\&=\sum_{e\in E(\dual)}\left(1-2\de_{e}(\la\oplus\eta)\right)=|E|-2|\la\oplus\eta|
\end{split}
\end{equation*}

In conclusion, we have
\begin{equation*}\label{}
\begin{split}
\sqrt{\frac{o(\z_{1})}{o(\z_{2})}} f_{\text{FK}}(\z_{1},\z_{2})&=\sum_{\om}\rho_{p}(\om){\bf 1}_{\ga:\z_{1}\to\z_{2}}e^{-\frac i2 {\bf w}(\ga(\om): \z_{1}\to \z_{2})} \\
&=\sum_{\om}\sum_{\s}\mu_{\la}(\s,\om){\bf 1}_{u_{1}\to u_{2}}e^{-\frac i2 {\bf w}(\ga(\om): \z_{1}\to \z_{2})} \\
&=\sum_{\om}\sum_{\s}\mu_{\la}(\s,\om){\bf 1}_{u_{1}\to u_{2}}i e^{-\frac i2 {\bf W}(\ga^{\text w}(\om): \z_{1}\to \z_{2})} \\
&=i \sum_{\om}\sum_{\s}\mu_{\la}(\s,\om) \s_{u_{1}}\s_{u_{2}}e^{-\frac i2 {\bf W}(\la: \z_{1}\to \z_{2})} \\
&=i \sum_{\s}\pi_{\beta}(\s)e^{-2\beta E_{\la}[\s]}\s_{u_{1}}\s_{u_{2}}e^{-\frac i2 {\bf W}(\la: \z_{1}\to \z_{2})} \\
&={i\over\sZ}\sum_{\la\oplus\eta\in\sC(\z_{1},\z_{2})}e^{-2\beta|\la\oplus\eta|}e^{-\frac i2 {\bf W}(\la\oplus\eta: \z_{1}\to \z_{2})}= \sqrt{\frac{o(\z_{1})}{o(\z_{2})}} f_{\text{Ising}}(\z_{1},\z_{2})
\end{split}
\end{equation*}
\endproof

\section{Many-point fermionic observable}\label{sec:many-point}
The natural generalization of the two-point fermionic observable to the $2n$-point consists in averaging loops winding over loop configurations where the $2n$ insertion points are pairwise connected by a loop. 
In this section we introduce this (well-posed) generalization for $2n$ insertion points and show that it coincides with the $2n$ point Ising observable defined in \cite{honglerthesis}.


In the definition of our observable, the {\it admissible configurations} are those in which each loop contains an even number of insertion points. The event $A(\om)$ indicates that the configuration $\om$ is admissible. As such, the FK-Ising fermionic observable with an odd number of insertion points, with free boundary conditions is by default null.
For an admissible configuration $\om$ there might be several ways of pairwise matching insertion points that are in the same loop. 

\begin{definition}[Perfect matching]
A perfect matching $\s$ of $2n$ points $\z_{1},\dots,\z_{2n}\in\Om$ is a pair partition $\{\{i_{j},\tau_{j}\}:j\in\{1,\dots,n\}\}$ of the indices $\{1,\dots,2n\}$ with $i_{j}<\tau_{j}, \ \ \forall \ j$. The sign of the perfect matching, $sign(\s)$ is defined as the the sign of the permutation $\mu_{\s}\in S_{2n}$ defined by $\mu_{\s}(2k-1)=i_{k}$ and $\mu_{\s}(2k)=\tau_{k}$ for $k=1,\dots,n$. We indicate with $\sC_{2n}$ the set of all possible perfect matchings.
\end{definition}

Every admissible configuration $\om$ such that the points $\z_{1},\dots,\z_{2n}$ are connected pairwise, induces a perfect matching $\s$. Such an induced perfect matching $\s$ is not necessarily unique, in particular in case in the configuration $\om$ some loops contain more than one pair of insertion points, then one has more choice for $\s$. However, when such a situation arise, a preferred matching can be chosen by exploring all the loops where the ambiguity arise and selecting a particular matching within the insertion points in each of those loops: one starts from the corner with lowest insertion index $i\in\{1,\dots,2n\}$ and pairs it with the next corner point encountered when walking the loop, thus one keeps walking and pairs together the next two encountered corner points, until exhaustion. Such an algorithm generates a matching $\tau$ called {\it sequential perfect matching}.


Therefore, for an admissible configuration $\om$, and its unique sequential perfect matching $\tau=\{\{i_{k},\tau_{k}\} : k\in\{1,\dots,n\}\}$, there exist $n$ paths $\ga_{k}$, $k\in\{1,\dots, n\}$ connecting $\z_{i_{k}}$ and $\z_{\tau_{k}}$ (either from $\z_{i_{k}}$ to $\z_{\tau_{k}}$ or in reversed order).
%
%

 \begin{definition}[$2n$-point fermionic observable]
Let $\z_{1},\dots,\z_{2n}\in\corner$ be distinct corner points. The {\it real} version of the $2n$-point fermionic observable is defined as
$$f_\text{FK}(\z_{1},\dots,\z_{2n})=\E\left[(-1)^{\s}{\bf 1}_{A}\phi(\s,\z_{1},\dots,\z_{2n})\right]$$
where $A(\om)$ is the event that $\om$ is an admissible configuration, the factor
$$\phi(\s;\z_{1},\dots,\z_{2n})(\om)=\prod_{k=1}^{n}\phi(\ga_{k}(\om),\z_{i_{k}},\z_{\tau_{k}})$$ is the total winding, viz. the product of the winding of all the pairwise connecting paths, walked having primal clusters on the left, in the configuration $\om$ determined by the unique sequential perfect matching $\s(\om)$.
\end{definition}

As a consequence of the composition property of the winding, Proposition \ref{prop:winding-properties}, one can notice that for the admissible configurations $\om$ in which a loop passes through four or more points, the possible different pairings within that loop have all the same winding.

%

Recall that we use the notation $\z_{i}$ to indicate the $i$-th entry in the observable $f(\z_{1},\dots,\z_{2n})$ and the notation $\z^{i}$ to indicate the $i$-th insertion point visited on a loop exploration.

\begin{proposition}\label{prop:adding-winding}
Let $\z^{1}\in\corner$ an insertion point on a loop $\ga$ and let $\z^{2},\dots,\z^{2m}$ the other insertion points on the loop $\ga$, indexed in the order of appearance when walking $\ga$ starting from $\z^{1}$. Then, for every perfect matching $\s=\{\{i_{j},\tau_{j}\} : j\in\{1,\dots,m\}\}$ of $\z^{1},\dots,\z^{2m}$ one has
$$(-1)^{\s}\prod^{m}_{j=1}\phi(\ga,\z^{i_{j}},\z^{\tau_{j}})=\prod^{m}_{j=1}\phi(\ga,\z^{2j-1},\z^{2j})$$
\end{proposition}
\proof
Recall that, by definition of perfect matching, $\z^{i_{1}}$ is the corner point with lowest entry index.
For each winding $\phi(\ga,\z^{i_{j}},\z^{\tau_{j}})$ one has $$\phi(\ga,\z^{i_{j}},\z^{\tau_{j}})=\prod_{u=0}^{\tau_{j}-i_{j}-1}\phi(\ga,\z^{i_{j}+u},\z^{i_{j}+u+1})$$
thus, by Proposition \ref{prop:winding-properties}, we have
\begin{equation*}\label{}
\begin{split}
(-1)^{\s}\prod^{m}_{j=1}\phi(\ga,\z^{i_{j}},\z^{\tau_{j}})&=(-1)^{\s}\prod_{j=1}^{m}\prod_{u_{j}=0}^{\tau_{j}-i_{j}-1}\phi(\ga,\z^{i_{j}+u_{j}},\z^{i_{j}+u_{j}+1})\\
&=\prod^{m}_{j=1}\phi(\ga,\z^{2j-1},\z^{2j})\prod^{m-1}_{j=1}\left(\phi(\ga,\z^{2j},\z^{2j+1})\right)^{2p_{j}}\\
&=\prod^{m}_{j=1}\phi(\ga,\z^{2j-1},\z^{2j})
\end{split}
\end{equation*}
The last equality follows by considering that segments $\ga:\z_{2j}\to\z_{2j+1}$ are walked by an even number of paths $2p_{j}$, $p_{j}\in\N$, while segments $\ga:\z_{2j-1}\to\z_{j}$ are walked by an odd number of paths $2p_{j}+1$. Regardless of the exact values of $p_{j}$, which depend on the particular choice of $\s$, the proof is conclude by recalling that $\phi^{2}(\ga)=1$. 
\endproof

The Proposition above shows that for an admissible configuration $\om$ and a perfect matching $\s\in\sC_{2n}$, with a loop going through several number of points, while a priori the total winding phase depends on the particular pairing selected by a $\s$, in truth the total winding phase can be computed as the product of the winding phases of the several paths connecting the insertion points sequentially and without intersection.

For a fixed admissible configuration $\om$, it is evident that the winding phase $\phi(\ga,\z^{1},\z^{2})$ does not actually depend on the whole loop $\Ga$ of $\om$ where the corner points lay on, but rather only on the path, portion of the loop, $\ga\subset \Ga$ from $\z^{1}$ to $\z^{2}$. In particular, if we modify the configuration $\om$, $\phi(\ga,\z^{1},\z^{2})$ stays constant as far as we do not modify the path $\ga$; eventually, even the portion of the loop $\Ga\setminus\ga$ can change without affecting $\phi(\ga,\z^{1},\z^{2})$.
Similarly, if one considers an admissible configuration, with a sequential perfect matching $\s$, with two non-intersecting paths $\ga_{1}$ from $\z^{1}$ to $\z^{2}$ and $\ga_{2}$ from $\z^{3}$ to $\z^{4}$, then any configuration that does not modify the paths $\ga_{1}$ and $\ga_{2}$ would give rise to the same winding $\phi(\ga_{1},\z^{1},\z^{2})$, and $\phi(\ga_{2},\z^{3},\z^{4})$. This means that in particular, whether $\z^{1},\z^{2},\z^{3},\z^{4}$ lie all on the same loop or in two distinct ones, as far as the paths $\ga_{1}$ from $\z^{1}$ to $\z^{2}$ and $\ga_{2}$ from $\z^{3}$ to $\z^{4}$ are the same, the winding phases coincide as well. Different configuration with same collections of paths $\{\ga_{i}\}$ are called {\it $\{\ga_{i}\}$-similar}. This means that for any admissible configuration $\om$ the total winding $\phi(\s;\z_{1},\dots,\z_{2n})(\om)$ is equal to the winding of a $\{\ga_{i}\}_{i=1:n}$-similar configuration consisting of only one loop connecting all the $2n$ points. Furthermore, thanks to \ref{prop:adding-winding} this is also equal to the product  $(-1)^{\s}\prod^{n}_{j=1}\phi(\ga,\z^{i_{j}},\z^{\tau_{j}})$ for any perfect matching $\s\in\sC_{2n}$.

\begin{proposition}\label{antisymmetry}
For any permutation $\s:\{1,\dots,2n\}\to\{1,\dots,2n\}$ one has
$$f_\emph{FK}(\z_{\s(1)},\dots,\z_{\s(2n)})=(-1)^{\s}f_\emph{FK}(\z_{1},\dots,\z_{2n})$$
\end{proposition}
\proof
It suffices to prove it for a transposition $\s_{j}$ that swaps the indices $j$ and $j+1\in\{1,\dots,2n\}$ leaving the others unchanged, for such a permutation we have $(-1)^{\s_{j}}=-1$. The set of admissible configuration stays unchanged and we only need to see how, for each configuration $\om$ and associated sequential perfect matching $\s(\om)$, the phase factor
$$(-1)^\s\phi(\s;\z_{1},\dots,\z_{2n})(\om)$$
changes. 

The transposition $\s_{j}$ might influence the phase in three ways:
\begin{itemize}
\item if both $j$ and $j+1$ are end-point of paths, i.e. $\exists k, k'$ s.t. $\tau_{k}=j$ and $\tau_{k'}=j+1$, then applying $\s_{j}$ let $\phi$ constant, but  changes the parity of crossing of $\s$, i.e. $\sign(\s_{j}\s)=-\sign(\s)$; and similarly for the case with both $j$ and $j+1$ being starting-point, i..e  $\exists k, k'$ s.t. $i_{k}=j$ and $i_{k'}=j+1$;
\item if $\exists k$ such that $j=i_{k}$ and $j+1=\tau_{k}$, or vice versa, then the perfect matching does not change $\s_{j}\s=\s$, the winding $\phi$, thanks to Propositions \ref{prop:winding-properties} and \ref{prop:adding-winding} gains a $-1$ sign.
\end{itemize}
\endproof

\subsection{Fermionic observable with generic boundary conditions.}\label{boundary conditions}
So far we have discussed the definition of $f$ for the FK model with {\it free} boundary conditions on the primal lattice -- i.e. fully wired boundary conditions on the dual lattice --. Such boundary condiitions constrain the configurations $\om$ to consists of loops only and no open path. The same occurs in the case of {\it fully wired} boundary conditions, for which the definition of the observable still holds, as it is invariant under exchanging the role of primal and dual lattice.

In the case of {\it mixed} boundary conditions, the boundary consists of alternating dual wired (free) and primal wired boundary arcs that connect boundary corners $a_{1},\dots,a_{2n}$. Such condition constrains the configurations to consist of a collection of loops and $n$ simple, non-inersecting paths connecting pairwise the boundary corners $a_{1},\dots,a_{2n}$. With these boundary conditions the definition of the observable does not change but for the definition of admissible configuration $\om$ (for which the event $A(\om)$ occur): for a fermionic observable with insertion points $\z_{1},\dots,\z_{2n}$, a configuration $\om$ is admissible if and only if each loop {\it and path} in $\om$ contains an even number of insertion points. 
Futhermore, each path from the boundary corner $a_{i}$ to the boundary corner $a_{j}$ can be topologically prolonged out of the planar domain $\Om$ from $a_{j}$ to $a_{i}$ so to form a simple loop. Consequently, Proposition \ref{prop:winding-properties}, and related lemmas, is valid for mixed boundary conditions too.

\subsection{Equivalence between FK-Ising and Ising $2n$-point observables}
The $2n$-point discrete fermionic observable for the Ising model has been defined in \cite{honglerthesis,hongler2013energy,hongler2013conformal,gheissari2019ising} by extending the two-point definition to the case in which $n$ corner defect lines are present. We recall here the definition of the observable as in \cite{hongler2013conformal}.

\begin{definition}[Ising $2n$-point fermionic observable]
Let $\z_{1}\dots\z_{2n}$ be distinct corners. Let $\La=\{\la_{1}:\z_{1}\to\z_{2},\dots,\la_{n}:\z_{2n-1}\to\z_{2n}\}$ be a collection of $n$ disjoint corner defect lines. We define the $2n$-point fermionic observable for the Ising model as
$$f_\text{Ising}(\z_{1},\dots,\z_{2n}):=\bra\prod_{j=1}^{n}(\psi(\z_{2j-1}),\psi(\z_{2j}))_{\la_{j}}\ket .$$
\end{definition}

As for the two-point case the equivalence between the two discrete fermions relies on the Edwards-Sokal coupling between the Ising and FK-Ising models and on the possibility of relating the winding phases defining the two functions.

\begin{theorem}\label{theorem:multipts-equivalence}
Let $\z_{1},\dots,\z_{2n}\in\corner$ be distinct corner points. Then the discrete fermionic observables for the FK-Ising model and for the Ising model defined on $\z_{1},\dots,\z_{2n}$ coincide.
\end{theorem}
\proof
Consider an admissible configuration $\om$, we have seen in Proposition \ref{prop:adding-winding} that starting from the collection of loops of $\om$ we can virtually modify the loops to a unique loop containing all the corner points, without changing the winding. Thus we can equate that winding to the winding of any other pair matching, up to the sign of the pair matching. In particular we can choose the same pair matching $
\alpha$ occurring for the collection of $n$ disjoint corner defect lines $\La=\{\la_{i}\}_{i=1:n}$. As the FK loop is non intersecting, pair by pair we can relate the FK-loop winding to the winding of the defect lines. Finally, by using the Edwards-Sokal coupling with $n$ disjoint defect line
$$\mu_{\La}(\s,\om)\propto\prod_{e\in E(\Om)\setminus\La}\left((1-p)\de_{\om(e),0}+p\de_{\om(e),1}\de_{e}(\s)\right)\prod_{e\in \La}\left((1-p)\de_{\om(e),0}+p\de_{\om(e),1}(1-\de_{e}(\s))\right)$$
for which we have

$$\sum_{\om\in\{0,1\}^{E(\Om)}}\mu_{\La}(\s,\om)=e^{-2\beta E_{\La}[\s]}\pi_{\beta}(\s) ; \qquad\sum_{\s\in\{\pm1\}^{V(\Om)}}\mu_{\La}(\s,\om)= \rho_{p}(\om)\prod_{k=1}^{n}{\bf 1}_{w_{i^{\alpha}_{k}}\to w_{\tau^{\alpha}_{k}}} \ .$$

Trivially, $$\phi(\ga_{k}(\om),\z_{i_{k}},\z_{\tau_{k}})\phi(\ga_{k}(\om),\z_{\tau_{k}},\z_{i_{k}})=-1 \ ,$$ and thus, for the total winding of any admissible FK configuration, as fixed by its sequential perfect matching, one has 
\begin{equation*}\label{}
\begin{split}
(-1)^{\s}\phi(\s;\z_{1},\dots,\z_{2n})(\om)=\prod_{k=1}^{n}\phi(\ga_{k}(\tilde \om),\z^{2k-1},\z^{2k})=(-1)^{\alpha}\prod_{k=1}^{n}\phi(\ga_{k}(\tilde \om),\z_{i^{\alpha}_{k}},\z_{\tau^{\alpha}_{k}})
\end{split}
\end{equation*}
for some $\{\ga_{i}\}$-similar configuration $\tilde\om$ where all the points $\z_{i}$ lay on the same loop.
Thus one introduce an Edwards-Sokal coupling in the presence of $n$ disjoint disorder lines $\la_{1},\dots,\la_{n}$.
\begin{equation*}\label{}
\begin{split}
f_{\text{FK}}(\z_{1},\dots,\z_{2n})&=\sum_{\om}\rho_{p}(\om)(-1)^{\s(\om)}{\bf 1}_{A}(\om)\phi(\s;\z_{1},\dots,\z_{2n})(\om) \\
&=\sum_{\om}\rho_{p}(\om){\bf 1}_{A}(\om)(-1)^{\alpha}\prod_{k=1}^{n}\sqrt{\frac{o(\z_{2k-1})}{o(\z_{2k})}}\exp\(-\frac i2 {\bf w}(\ga_{k}(\tilde\om): \z_{i^{\alpha}_{k}}\to \z_{\tau^{\alpha}_{k}})\) \\
&=i^{n}\sum_{\om}\sum_{\s}\mu_{\La}(\s,\om)\prod_{k=1}^{n}\s_{u_{2k-1}}\s_{u_{2k}}\sqrt{\frac{o(\z_{2k-1})}{o(\z_{2k})}}\exp\(-\frac i2 {\bf W}(\la_{k}: \z_{2k-1}\to \z_{2k})\) \\
&=\sum_{\s}\pi_{\beta}(\s)\prod_{k=1}^{n}(\psi(\z_{2k-1}),\psi(\z_{2k}))_{\la_{k}}=f_{\text{Ising}}(\z_{1},\dots,\z_{2n})
\end{split}
\end{equation*}
\endproof

\subsec{FK Exploration tree}
The loops of an FK configuration can be visited via an exploration tree. The 2n-point discrete fermionic observable can thus be equivalently defined in terms of the exploration tree. This approach will be particularly convenient for the study of the observable in the continuous limit $\de\to0$, see \ref{sec:beyond}.

\begin{definition}[FK-Exploration branching tree]
Given a FK-loop configuration with fully wired boundary conditions $\om$ and a point on the boundary $a\in\partial\Om$, we define an FK-exploration branching tree with the following procedure. The tree $(\om,a)\mapsto T$ is the branching binary tree obtained with the following exploration process: \begin{itemize}
\item (exploration) starting from $a$ cut open the loop next to $a$ and walk the loop keeping primal edges on the left;
\item (branching) when the path arrives at a point which disconnect the domains in two subdomains, we have a {\it branching point} $a_{t}$ for the exploration tree: a branch of the tree will proceed on the same loop;
\item (recursion) the other branch explored the to-be-disconnected domain: one cut open the loop next to $a_{t}$ and proceed again with the exploration in the subdomain.
\item The process stops when the whole domain has been explored.
\end{itemize}
\end{definition}

\begin{figure}
    \includegraphics[width=.5\linewidth]{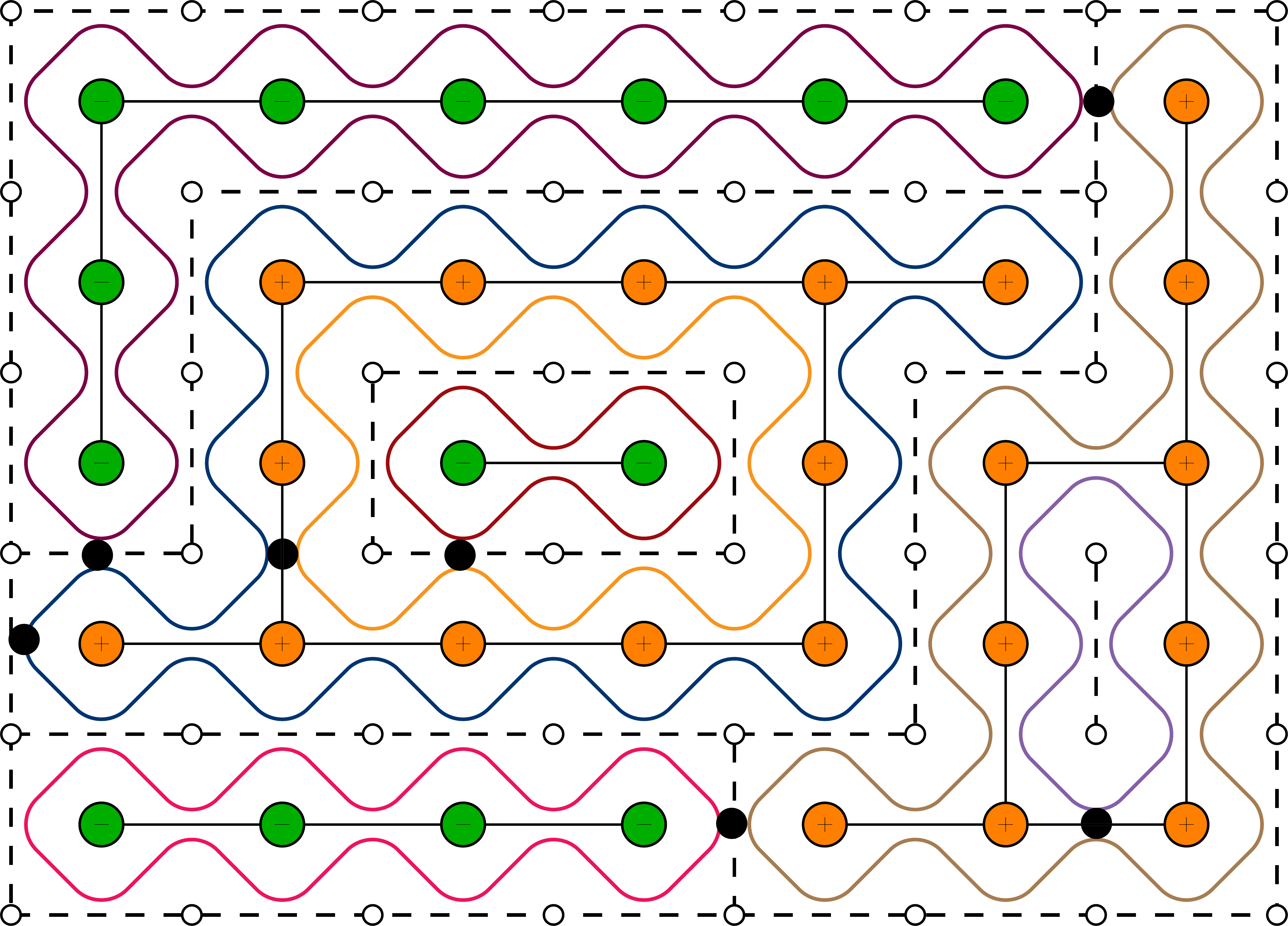}
    \caption{\label{fig:exploration} Exploration tree for FK loop configuration of Fig. \ref{fig:loop-config}. The exploration starts at black point on the boundary and it explores counterclockwise the (blue) loop until it disconnects into two domains (black dots) and it branches: one branch keeps exploring the blue loop, the other one starts exploring the yellow loop. The exploration proceed until all the loops are explored.}
\end{figure}


\begin{definition}
For a given branch $u\in T(\om,a)$ and $2n$ points $\z_{1},\dots,\z_{2n}$	 the winding of the branch $W(u)$  is defined as 
$$W(u)={\bf1}_{P(u)}\prod_{\z_{\s(i)}\in u,\  i\equiv_{2}1}\phi(u,\z_{\s(i)},\z_{\s(i+1)})(-1)^{{\bf 1}_{\s(i)>\s(i+1)}} .$$
\end{definition}

Intuitively, when a branch $u$ cross an insertion point $\z$ a clock starts that measure the winding; when $u$ meets a second insertion point $\z'$ the clock is stopped. If $u$ meets overall an odd number of insertion points the loop configuration is not an admissible and it is discarded. One has $\prod_{u\in T}{\bf 1}_{P(u(\om))}={\bf 1}_{A(\om)}$. Furthermore the total winding $W(T(\om,a)):=\prod_{u\in T(\om,a)}W(u)$ does not depend on the choice of the initial point $a$ and the branching points and it is immediate to verify that $\E[W(T)]=f_{\text{FK}}(\z_{1},\dots,z_{2n})$.

\sec{Discrete holomorphicity}
In Theorems \ref{theorem:2pts-equivalence} and \ref{theorem:multipts-equivalence}, in order to show the equivalence between the discrete fermionic observable in the Ising model and in the FK-Ising model, we established the connection between FK loops going through the insertion points and disorder lines. Such a proof show the equivalence of the two observables for any value of $p=1-e^{-2\beta}$.
However, when the parameters are tuned to their critical values, $$p_{c}={\sqrt2\over 1+\sqrt2},  \qquad\beta_{c}=\frac12\ln(\sqrt2+1),$$ there is an an alternative proof, based on the {\it (strong) discrete holomorphicity} (also known as s-holomorphic \cite{smirnov2010conformal}) of the observables: one formulates a discrete {\it Riemann Boundary Value Problem} that admits a unique s-holomorphic solution, and shows that both observables are solutions of such a problem.

Furthermore, s-holomorphicity can be used to reveal the pfaffian structure of the discrete fermionic observable \cite{honglerthesis},

$$f(\z_{1},\dots,\z_{2n})=\pfaffian({\bf F}(\z_{1},\dots,\z_{2n}))$$

where the antisymmetric matrix ${\bf F}(\z_{1},\dots,\z_{2n})\in{\bb M}_{2n}(\bb R)$ has entries 

\begin{equation*}\label{}
({\bf F}(\z_{1},\dots,\z_{2n}))_{jk}=
\begin{cases}
f(\z_{j},\z_{k}) \qquad j\neq k \\
0 \qquad\qquad\quad \text{otherwise}
\end{cases}
\end{equation*}

Such a structure shed a light on the free fermion nature in the Ising theory already at the discrete level. On this direction, construction of a lattice Virasoro algebra for the Ising model has been carried out completely in \cite{hongler2013conformal}.

Since strong holomorphicity only holds at criticality, a proof based on it is weaker than the one proposed above, however, by easily yielding precompactness estimates, it plays a pivotal role in proving convergence of the observables to correlation functions of the continuous fermion and in proving convergence of FK$_{2}$ loops to $\cle_{16\over3}$ \cite{duminil2012conformal,chelkak2014convergence,garban2018convergence}.

Away from criticality, discrete analyticity persists only in a perturbed sense \cite{hongler2014discrete,duminil2014near}, and gives rise in a {\it massive scaling limit} to the convergence of the discrete Ising fermion to a {\it massive fermion} \cite{park2018massive}. The equivalence between FK-Ising and Ising discrete observable extends then results for the Ising model to the FK model too.

In the following sections we show strong holomorphicity for the fermionic observable and its pfaffian structure. The equivalence of the FK-Ising observable and the Ising observable follows then from \cite{hongler2013conformal}, by observing that the difference of the two functions is s-holomorphic everywhere, and thus constant, and it attains the value zero. 

\subsec{Strong holomorphicity}
Strong holomorphicity of the discrete fermion is the discrete precursor of the holomorphicity (chirality) of the Ising free fermion and it is a property of functions defined on mid-edges $z\in\midpoint$.
\begin{definition}[Strong holomorphicity]
A function $f:\midpoint\to\C$ is \emph{strong holomorphic}, or \emph{s-holomorphic} if for each pair of adjacent midpoints $z_{1},z_{2}\in\midpoint$ with a common corner point $\z$ one has
$$\bf{P}_{(io(\z))^{-\frac12}}[f(z_{1})]=\bf{P}_{(io(\z))^{-\frac12}}[f(z_{2})] .$$
\end{definition}
where $\bf{P}_{\nu}[X]=\Re\left(X{\bar \nu\over |\nu|}\right)$ indicates the orthogonal projection of $X$ on $\nu$.

If a function is strong holomorphic then it is also discrete holomorphic \cite{chelkak2011discrete}, in the sense that the discrete contour integral of its projection around a mid-edge $z$ is null
\begin{equation}\label{cauchy}
\begin{split}
\sum_{\sC(z)}\bf{P}_{(io(\z))^{-\frac12}}[f(z_{1})] d\z =0 \ .
\end{split}
\end{equation}
where $\sC(z)$ is any simple path along corners enclosing $z$.
The reader recognizes in such a property the discrete version of Cauchy's integral theorem.

Strong holomorphicity is a property of functions of one variable midpoint, so in order to state strong holomorphicity results for the discrete fermions $f(\z_{1},\dots,\z_{2n})$ --throught the section we omit the FK indication from $f$-- we consider all the entries $\z_{1},\dots,\z_{2n-1}\in\corner$ fixed and extend $f$ from a function defined on corners $\z=\z_{2n}\in\corner$ distinct from $\z_{1},\dots,\z_{2n-1}$ to a function defined on mid-edges $\midpoint\ni z\longmapsto F(\z_{1},\dots,\z_{2n-1},z)$. In addition, we assume that $\z_{1},\dots,\z_{2n-1}$ are corners that pairwise do not share any primal or dual adjacent vertex.

In order to extend $\z\mapsto f(\z_{1},\dots,\z_{2n-1},\z)$ to mid-edges, we notice that for any inner mid-edge $z\in\midpoint$ the four adjacent corners $z_\text{NW},z_\text{NE},z_\text{SE},z_\text{SW}\in\corner$ satisfy the identity
$$\bf{1}_{z_\text{NW}\in\ga}+\bf{1}_{z_\text{SE}\in\ga}=\bf{1}_{z_\text{SW}\in\ga}+\bf{1}_{z_\text{NE}\in\ga}$$
Furthermore, we notice that for two adjacent corners, the turning $\rho$ from one corner $\z_{a}$ to the other $\z_{b}$, keeping the primal vertex on its left is such that $\phi(\rho:\z_{a}\to\z_{b})=1$. Thus for any FK configuration $\om$, if the path $\ga$ connects a corner $\z\in\corner$ to the corner $z_\text{SE}$ (and we recall that paths are walked keeping primal cluster on their left) then it necessarily goes first through either $z_\text{NE}$ or $z_\text{SW}$, thus we either $\phi(\ga:\z\to z_\text{SE})=\phi(\ga:\z\to z_\text{NE})$ or  $\phi(\ga:\z\to z_\text{SE})=\phi(\ga:\z\to z_\text{SW})$, and similarly for $z_\text{NW}$.
This means that if $z\in\midpoint$ is a mid-edge not adjacent to any $\z_{1},\dots,\z_{2n-1}$, and the system is at criticality, i.e. when the measure $\rho_{p}(\om)\propto \sqrt{2}^{\ell(\om)}$, then
\begin{equation}\label{nice equation}
\begin{split}
f(\z_{1},\dots,\z_{2n-1},z_\text{NW})+f(\z_{1},\dots,\z_{2n-1},z_\text{SE})=f(\z_{1},\dots,\z_{2n-1},z_\text{NE})+f(\z_{1},\dots,\z_{2n-1},z_\text{SW})
\end{split}
\end{equation}
Notice that if $t\neq1$, whether the primal edge through $z$ is open or closed will affect the values of $f$, and the identity \eqref{nice equation} would not hold \cite{riva2006holomorphic,alam2014integrability}.

From now on, we fix $t=1$ and work in the context of the critical FK-Ising model only.
We can then extend the fermionic observable on mid-edge as
\begin{equation*}\label{}
\begin{split}
\midpoint\ni z\longmapsto F(\z_{1},\dots,\z_{2n-1},z)&=f(\z_{1},\dots,\z_{2n-1},z_\text{NW})+f(\z_{1},\dots,\z_{2n-1},z_\text{SE})\\
&=f(\z_{1},\dots,\z_{2n-1},z_\text{NE})+f(\z_{1},\dots,\z_{2n-1},z_\text{SW})
\end{split}
\end{equation*}
In the case where $z$ is a mid-edge adjacent to any of $\z_{1},\dots,\z_{2n-1}$, equation
\eqref{nice equation} does not hold anymore - as one of the four values of $f$ is not defined. 

Away from the diagonals $\z=\z_{i}$ we have that $f$, or to better say a rotate version of $f$, is strong holomorphic.

\begin{proposition}
Let $\z_{1},\dots,\z_{2n-1},\z\in\corner$ be distinct corners, and let $f$ be the FK 2n-point fermionic observable. Let
$$h(\z_{1},\dots,\z_{2n-1},\z):=\sqrt{-i \over o(\z)}f(\z_{1},\dots,\z_{2n-1},\z) \ .$$
Then, for every $z\in\midpoint$ mid-edge not adjacent to any $\z_{i}$, $i=1,\dots,2n-1$ the function
$$H(\z_{1},\dots,\z_{2n-1},z):=h(\z_{1},\dots,\z_{2n-1},\z) +h(\z_{1},\dots,\z_{2n-1},\z^{o}) .$$
is s-holomorphic.
\end{proposition}
\proof
While $h(\z_{1},\dots,\z_{2n-1},\z)$ is parallel to the projection line $\sqrt{1\over o(\z)}$, $h(\z_{1},\dots,\z_{2n-1},\z^{o})$ is orthogonal, since $o(\z)$ and $o(\z^{o})$ are opposite in direction.
\endproof

\subsec{Discrete residue calculus}
The function $H(\z_{1},\dots,\z_{2n-1},z)$ is s-holomorphic on all midpoints $\z\in\midpoint$ not adjacent to $\z_{1},\dots,\z_{2n-1}$. These corner values correspond to discrete singularities for $H$ (and thus for $f$) and as such the discrete Cauchy integral in \eqref{cauchy} is non-zero for non contractible simple paths in $\Omega\setminus\z_{1},\dots,\z_{2n-1}$ and the integrand will have a discrete residue.

We have seen that if $z\in\midpoint$ is a mid-edge adjacent to any $\z_{1},\dots,\z_{2n-1}$, the identity \eqref{nice equation} does not hold because $f(\z_{1},\dots,\z_{2n-1},\z_{i})$ is not defined. However, since the values of $f(\z_{1},\dots,\z_{2n-1},\z)$ with $\z$ being any of the other three corners in the same placquette of $\z_{i}$ are well defined, we can extend the definition of $f$ along the diagonal by imposing eq. \eqref{nice equation} to hold. Let us suppose that $z$ is adjacent to $\z_{i}$, and let us use the notation $\z_{i}^{o}$ for the corner symmetric to $\z_{i}$ with respect to $z$, $\z_{i}^{l}$ for the corner on the left of $\z^{i}$, and $\z_{i}^{r}$ for the corner on its right.
\begin{equation}\label{fpm}
\begin{split}
f(\z_{1},\dots,\z_{2n-1},\z_i)=f(\z_{1},\dots,\z_{2n-1},\z_i^{r})+f(\z_{1},\dots,\z_{2n-1},\z_i^{l})-f(\z_{1},\dots,\z_{2n-1},\z_i^{o})
\end{split}
\end{equation}
However, any corner $\z\in\corner$ is adjacent to two different mid-edges $z_{1},z_{2}\in\midpoint$ so along the diagonals $\z=\z_{i}$ the function $f$ can be extended in two ways --let us use the notation $f^{+}$ and $f^{-}$ to indicate these two  values--. 

In the context of discrete complex analysis, the difference between the two values, $f^{+}-f^{-}$, corresponds to the residue of a discrete pole located in $\z_{i}$, i.e. to the value of the integral in \eqref{cauchy} when the path $\sC(z)$ is a simple path that surrounds only the singularity $\z_{i}$.

\begin{lemma}\label{two-point lemma}
For the two-point fermionic observable one has
$$|f^{+}(\z,\z)-f^{-}(\z,\z)|=2$$
\end{lemma}
\proof
First of all one has that $$\bbP[{\z\to\z^{l}}]+\bbP[{\z\to\z^{r}}]-\bbP[{\z\to\z^{o}}]=1 ;$$ and $$\phi(\z\to\z^{r})=\phi(\z\to\z^{l}) .$$
Furthermore, any path that goes from $\z$ to $\z^{o}$, right after $\z^{o}$, either encounters $\z^{l}$ or $\z^{r}$, depending on it one has either $\phi(\z\to\z^{o})=\phi(\z\to\z^{l})$ or $\phi(\z\to\z^{o})=\phi(\z\to\z^{r})$. So, the winding is the same, regardless of any ending point. Thus one has
$$f(\z,\z)=f(\z,\z^{l})+f(\z,\z^{r})-f(\z,\z^{o})=\phi(\z\to\z^{r}) .$$
The value of $\phi(\z\to\z^{r})$ is equal to $1$ in absolute value but, since the orientation of the path changes --once is incoming to and once is outgoing from $\z$-- its sign changes for the two mid-edges.
\endproof

\begin{lemma}
Let $\z_{1},\dots,\z_{2n-1}\in\corner$ be distinct corners, and let $f^{\pm}$ defined as in \eqref{fpm}, then
$$f^{\pm}(\z_{1},\dots,\z_{2n-1},\z_{1})=f(\z_{2},\dots,\z_{2n-1})f^{\pm}(\z_{1},\z_{1})$$
\end{lemma}
\proof
Consider one of the two cases, say $f^{+}(\z_{1},\dots,\z_{2n-1},\z_{1})$ associate to one of the two mid-edges. For an FK configuration $\om$, $A^{l}(\om)$ indicates the event that $\om$ is an admissible configuration, i.e. that the corners $\z_{1},\dots,\z_{2n-1},\z^{l}_{1}$ are pairwise connected, and similarly for $\z^{r}_{1}$ and $\z^{o}_{1}$; $B(\om)$ indicates that the corners $\z_{2},\dots,\z_{2n-1}$ are pairwise connected.
We also indicate with $\tilde \s(\om)$ the sequential perfect matching between $\z_{2},\dots,\z_{2n-1}$.

Let $\ga$ be the loop going through $\z_{1}$. Since each loop has to contain an even number of insertion points, for any configurations $\om$ we have the two alternatives
\begin{itemize}
\item both $A^{l}(\om)$, $A^{r}(\om)$ (and thus also $A^{o}(\om)$) simultaneously occur:
\item either $A^{l}(\om)$ occurs and $A^{r}(\om)$ does not, or vice versa (and thus $A^{o}(\om)$ does not occur).
\end{itemize}
The first case corresponds to configuration in which $\z_{1},\z_{1}^{l},\z_{1}^{r},\z_{1}^{o}$ all belong to the same loop $\ga$,
$${\bf 1}_{A^{l}(\om)}{\bf 1}_{A^{r}(\om)}={\bf1}_{\z^{l}\in\ga}{\bf1}_{\z^{r}\in\ga}{\bf 1}_{B(\om)}$$
for these configuration we have already seen that
$$\phi(\s(\om),\z_{1},\dots,\z_{2n-1},\z_{1}^{o})=\phi(\s(\om),\z_{1},\dots,\z_{2n-1},\z_{1}^{l})=\phi(\s(\om),\z_{1},\dots,\z_{2n-1},\z_{1}^{r})$$
 while the second case correspond to configurations in which $\z_{1}$ and $\z_{1}^{o}$ belong to different loops
 $${\bf 1}_{A^{l}(\om)}(1-{\bf 1}_{A^{r}(\om)})+{\bf 1}_{A^{r}(\om)}(1-{\bf 1}_{A^{l}(\om)})={\bf1}_{\z^{l}\in\ga}(1-{\bf1}_{\z^{r}\in\ga}){\bf 1}_{B(\om)}+{\bf1}_{\z^{r}\in\ga}(1-{\bf1}_{\z^{l}\in\ga}){\bf 1}_{B(\om)} \ .$$
Configurations $\om$ for which ${\bf1}_{\z_{1}^{l}\in\ga}(1-{\bf1}_{\z_{1}^{r}\in\ga})=1$ have a path $\ga$ that goes necessarily through $\z_{1}$ and then through $\z_{1}^{l}$, so the sequential perfect matching $\s(\om)$ matches $\z_{1}$ with $\z_{1}^{l}$ so the contribution of the configuration to $f(\z_{1},\dots,\z_{2n-1},\z_{1}^{l})$ is
$$\phi(\z_{1}\to\z_{1}^{l})\rho_{p}(\om){\bf 1}_{B}(-1)^{\tilde\s(\om)}\phi(\s,\z_{2},\dots,\z_{2n-1}) .$$
Similarly, configurations $\om$ for which ${\bf1}_{\z_{1}^{r}\in\ga}(1-{\bf1}_{\z_{1}^{l}\in\ga})=1$ give a contribution to $f(\z_{1},\dots,\z_{2n-1},\z_{1}^{r})$ 
$$\phi(\z_{1}\to\z_{1}^{r})\rho_{p}(\om){\bf 1}_{B}(-1)^{\tilde\s(\om)}\phi(\s,\z_{2},\dots,\z_{2n-1}) .$$
Finally, for configurations $\om$ for which ${\bf 1}_{\z_{1}^{o}\in\ga}={\bf 1}_{\z_{1}^{l}\in\ga}{\bf 1}_{\z_{1}^{r}\in\ga}=1$, the sequential perfect matching $\s(\om)$ might not necessarily associate $\z_{1}$ to $\z_{1}^{o}$, nonetheless, thanks to Proposition \ref{prop:adding-winding} one still has a contribution of
$$\phi(\z_{1}\to\z_{1}^{o})\rho_{p}(\om){\bf 1}_{B}(-1)^{\tilde\s(\om)}\phi(\s,\z_{2},\dots,\z_{2n-1})\ .$$
As seen in Lemma \ref{two-point lemma},
$$f^{+}(\z_{1},\z_{1})=\phi(\z_{1}\to\z_{1}^{l})=\phi(\z_{1}\to\z_{1}^{r})=\phi(\z_{1}\to\z_{1}^{o}) \ .$$

Overall, we have
\begin{equation}\label{}
\begin{split}
f^{+}(\z_{1},\dots,\z_{2n-1},\z_{1})&=f(\z_{1},\dots,\z_{2n-1},\z_{1}^{l})+f(\z_{1},\dots,\z_{2n-1},\z_{1}^{r})-f^{+}(\z_{1},\dots,\z_{2n-1},\z_{1}^{o})\\
&=\E[(-1)^{\s}{\bf 1}_{A^{l}}\phi(\s,\z_{1},\dots,\z_{2n-1},\z_{1}^{l})]\\
&+\E[(-1)^{\s}{\bf 1}_{A^{r}}\phi(\s,\z_{1},\dots,\z_{2n-1},\z_{1}^{r})]\\
&-\E[(-1)^{\s}{\bf 1}_{A^{l}}{\bf 1}_{A^{r}}\phi(\s,\z_{1},\dots,\z_{2n-1},\z_{1}^{o})]\\
&=\E[(-1)^{\s}{\bf 1}_{A^{l}}(1-{\bf 1}_{A^{r}})\phi(\s,\z_{1},\dots,\z_{2n-1},\z_{1}^{l})]\\
&+\E[(-1)^{\s}{\bf 1}_{A^{r}}(1-{\bf 1}_{A^{l}})\phi(\s,\z_{1},\dots,\z_{2n-1},\z_{1}^{r})]\\
&+\E[(-1)^{\s}{\bf 1}_{A^{l}}{\bf 1}_{A^{r}}\phi(\s,\z_{1},\dots,\z_{2n-1},\z_{1}^{o})]\\
&=f^{+}(\z_{1},\z_{1})\E[(-1)^{\tilde\s}{\bf1}_{\z^{l}\in\ga}(1-{\bf1}_{\z^{r}\in\ga}){\bf 1}_{B(\om)}\phi(\tilde\s,\z_{2},\dots,\z_{2n-1})]\\
&+f^{+}(\z_{1},\z_{1})\E[(-1)^{\tilde\s}{\bf1}_{\z^{r}\in\ga}(1-{\bf1}_{\z^{l}\in\ga}){\bf 1}_{B(\om)}\phi(\tilde\s,\z_{2},\dots,\z_{2n-1})]\\
&+f^{+}(\z_{1},\z_{1})\E[(-1)^{\tilde\s}{\bf1}_{\z^{r}\in\ga}{\bf1}_{\z^{l}\in\ga}{\bf 1}_{B(\om)}\phi(\tilde\s,\z_{2},\dots,\z_{2n-1})]\\
&=f^{+}(\z_{1},\z_{1})\E[(-1)^{\tilde\s}{\bf 1}_{B(\om)}\phi(\tilde\s,\z_{2},\dots,\z_{2n-1})]=f(\z_{2},\dots,\z_{2n-1})f^{+}(\z_{1},\z_{1})\\
\end{split}
\end{equation}
For $f^{-}(\z_{1},\dots,\z_{2n-1},\z_{1})$ one proceeds similarly, but the value of the winding $\phi(\z_{1}\to\z^{l}_{1})$ will be opposite.
\endproof

\begin{proposition}\label{residue}
Let $\z_{1},\dots,\z_{2n-1}\in\corner$ be distinct corners, and let $f^{\pm}$ defined as in \eqref{fpm}, then
$$f^{\pm}(\z_{1},\dots,\z_{2n-1},\z_{j})=(-1)^{j+1}f(\z_{1},\dots,\check\z_{j},\dots,\z_{2n-1})f^{\pm}(\z_{j},\z_{j})$$
\end{proposition}
\proof
Follows directly from Proposition \ref{antisymmetry} and from the lemma above. 
\endproof

\subsec{Pfaffian structure}
We are now in position to show that the $2n$-point real fermionic observable can indeed be written as the pfaffian of the antisymmetric matrix having the two-point fermionic observables as entries.
\begin{theorem}
Let $\z_{1},\dots,\z_{2n}$ be distinct corners. Then we have
$$f(\z_{1},\dots,\z_{2n})=\pfaffian({\bf F}(\z_{1},\dots,\z_{2n}))$$

where the antisymmetric matrix ${\bf F}(\z_{1},\dots,\z_{2n})\in{\bb M}_{2n}(\bb R)$ is defined as

\begin{equation*}\label{}
({\bf F}(\z_{1},\dots,\z_{2n}))_{jk}=
\begin{cases}
f(\z_{j},\z_{k}) \qquad j\neq k \\
0 \qquad\qquad\quad \text{otherwise}
\end{cases}
\end{equation*}
\end{theorem}
\proof
By definition, we have
$$\pfaffian({\bf F}(\z_{1},\dots,\z_{2n}))=\sum_{j=1}^{2n-1}(-1)^{j+1}f(\z_{1},\dots,\check\z_{j},\dots,\z_{2n-1})f(\z_{j},\z_{2n}) .$$
Let us consider the function
$$r(\z_{1},\dots,\z_{2n-1},z):=F(\z_{1},\dots,\z_{2n-1},z)-\sum_{j=1}^{2n-1}(-1)^{j+1}f(\z_{1},\dots,\check\z_{j},\dots,\z_{2n-1})F(\z_{j},z)$$
Away form the corners $\z_{i}$ $r$ is a real linear combination of s-holomorphic fucntions, so it stays s-holomorphic. For $z$ close to $\z_{i}$, the value of the projection of the two orthogonal components is $0$, but thanks to Proposition \ref{residue} but also the value of the projection of the two parallel components is $0$, and so it is s-holomorphic.
Thus, $r$ is s-holomorphic on all mid-edges $z\in\midpoint$; one can then use the maximum principle for strong holomorphic observables \cite{chelkak2012universality}, to conclude that $r$ is everywhere null.
\endproof

\section{Ising free fermion in CLE$(\frac{16}{3})$}\label{sec:beyond}
In the scaling limit $\delta\to0$ of the FK-Ising model, while fermionic correlation functions are given by the Ising CFT, loops are described by $\cle\left({16\over3}\right)$, the Conformal Loop Ensemble measure \cite{sheffield2012conformal} with parameter $\kappa={16\over 3}$. As such, thanks to exact convergence results \cite{benoist2019scaling,garban2018convergence} of both discrete correlations and paths to their countinuum counterpart, the reader can expect that the loop interpretation of fermions still holds in the continuum.

In fact, in a subsequent note we will show this result: the $2n$-point correlation of fermions in a simply connected domain $\Omega$, $\bra \psi(z_{1})\dots\psi(z_{2n})\ket_{\Omega}$ are also described by a suitable complexification of the probability that each $\cle\left({16\over3}\right)$ loops in $\Omega$ contain (i.e. ``they are $\epsilon$-close to'') an even number of insertion points $z_{j}$.


The general strategy is based on running an exploration tree: like in the discrete case, the correlation function $\bra \psi(z_{1})\dots\psi(z_{2n})\ket_{\Omega}$ can be obtained by averaging the total winding over all possible $\cle({16\over 3})$ exploration trees. One can group the different exploration tree with respect to the order of visit of the insertion points: in the case of boundary insertion points, for a fixed order of visits, the observable reduces to the probability of visiting the points in that order, which can be computed via quantum group techniques \cite{kytola2006conformal,Kyt_l__2019}.
The winding for points not on the boundary is obtained by analytically extending boundary visits probabilities in the bulk; finally to show the agreement of the result with the correlation function of the Ising fermions computed via CFT, i.e. $$\bra \psi(z_{1})\dots\psi(z_{2n})\ket_{\Omega}=\pfaffian\left({\bf \Psi}(z_{1},\dots,z_{n})\right),$$ where ${\bf \Psi}(z_{1},\dots,z_{n})$ being the antisymmetric matrix with non-diagonal entries ${\bf \Psi}_{ij}=\bra \psi(z_{i})\psi(z_{j})\ket_{\Omega}$, it will suffice to show that the singular parts and the boundary conditions of the two sides of the equation coincide.

The technique itself is of course not limited to homogeneous boundary conditions but works as well for more general boundary conditions. For instance, for Dobrushin wired/free boundary conditions, the representation of the two-point Ising correlation function $\bra \psi(z)\psi(w)\ket_{\Omega}^{[x,y]}$ is proven in \cite{hongler2013ising}, and can be extend by the same argument to the many-point case.

\bibliographystyle{alpha}
\bibliography{June2019.bib}
\end{document}